\documentclass[a4paper,11pt]{article}
\pdfoutput=1 

\usepackage{jcappub} 

\usepackage[T1]{fontenc} 
\usepackage{caption}

\usepackage{siunitx} 
\usepackage[dvipsnames]{xcolor}
\usepackage{mathtools} 
\usepackage{textcomp} 
\usepackage{caption}
\usepackage{hyperref}
\usepackage{subcaption,graphicx}
\hypersetup{
    colorlinks=true,
    linkcolor=blue,
    filecolor=magenta,      
    urlcolor=blue,
}
\usepackage[sort&compress,numbers]{natbib}
\newcommand{\citeb}[1]{\textcolor[rgb]{0,0,1}{\citep{#1}}}

\usepackage{color}
\usepackage{xcolor,colortbl}
\definecolor{ao(english)}{rgb}{0.0, 0.5, 0.0}
\usepackage{cancel} 
\usepackage[toc,page]{appendix}

\newlength\tindent
\renewcommand{\indent}{\hspace*{\tindent}}
\definecolor{purple}{RGB}{219, 70, 134}
\definecolor{greenish}{RGB}{90, 219, 154}
\definecolor{pinkish}{RGB}{219, 120, 34}

\newcommand\JCH[1]{\textcolor{black}{#1}}

\newcommand\Correx[1]{\textcolor{black}{#1}}

\def\eqref#1{\textcolor{blue}{(\ref{#1})}} 

\newcommand{\beq}{\begin{equation}}
\newcommand{\eeq}{\end{equation}}
\newcommand{\bea}{\begin{eqnarray}}
\newcommand{\eea}{\end{eqnarray}}
\newcommand{\beqn}{\begin{equation*}}
\newcommand{\eeqn}{\end{equation*}}
\newcommand{\bean}{\begin{eqnarray*}}
\newcommand{\eean}{\end{eqnarray*}}

\newcommand*{\cref}[1]{Chapter~\ref{#1}}

\title{\boldmath Complex Scalar Field Reheating and Primordial Black Hole production}

\author[a]{Karim Carrion,}
\author[a]{Juan Carlos Hidalgo,}
\author[a]{Ariadna Montiel,}
\author[a,b]{and Luis E. Padilla}

\affiliation[a]{Instituto de Ciencias F\'{\i}sicas, Universidad Nacional Aut\'onoma de M\'exico,\\ Cuernavaca, Morelos, 62210, Mexico}
\affiliation[b]{Mesoamerican Centre for Theoretical Physics,
Universidad Aut\'onoma de Chiapas, Carretera Zapata Km. 4, Real
del Bosque (Ter\'an), Tuxtla Guti\'errez 29040, Chiapas, M\'exico}

\emailAdd{hidalgo@icf.unam.mx}
\emailAdd{kcarrion@icf.unam.mx}
\emailAdd{amontiel@icf.unam.mx}
\emailAdd{luis.padilla@unach.mx}

\abstract{

We study perturbations of a complex scalar field during reheating with no self-interaction in the regime $ \mu \gg H$, when the scalar field has a fast oscillatory behaviour (close to a pressure-less fluid). We focus on the precise determination of the instability scale \Correx{and find it differs from that associated with a real scalar field. We further look at the probability that unstable fluctuations form Primordial Black Holes (PBHs) obtaining a significant production of tiny PBHs which quickly evaporate and may subsequently leave a population of Planck-mass relics. We finally impose restrictions on the duration and energy scale of the fast oscillations period by considering that such relics constitute, at most, the totality of dark matter in the Universe.} \\ 

\begin{flushright}
\indent \textbf{Key words:} Cosmology, Primordial Black Holes, Scalar Field, Reheating.  
\end{flushright}

}

\begin{document}
\maketitle

\flushbottom

\section {\bf Introduction}
\label{Int}

\indent For decades, cosmologists have been interested in Scalar Fields (SFs) and their role in the evolution of our Universe \citeb{zee2010quantum,dodelson2003modern}. Their dynamics are usually described by the Einstein-Klein-Gordon (EKG) system of equations, which can be seen as a relativistic generalization of the Schr\"odinger-Poisson or the Gross-Pitaesvkii-Poisson system (with the second case arising when self-interaction is considered). This EKG system was originally considered in the context of boson stars \citeb{bs1,bs2,bs3,bs4,bs5,bs6,bs7,bs9,bs10,bs11,bs12,bs13,bs14}\footnote{Boson stars are graviational bound objects constructed of complex SFs. In the case in which the SF is real, no stationary configurations exist. Instead, the configurations are oscillatory solutions, reason why they are dubbed oscillatons \citeb{osc1,osc2,osc3,osc4}.}, motivated by the axion field -- a pseudo-Nambu-Goldstone boson of the Peccei-Quinn phase transition -- which was originally proposed as a solution to the strong CP problem in QCD. However, due to the rich dynamics that a SF can have in a cosmological context, it has been proposed to use them in different scenarios in cosmology, for instance: Dark Matter (Scalar Field Dark Matter, ``SFDM'', see e.g. \citeb{Matos:2000ng,Matos:1998vk,Matos:1999et,Peebles:1999se,Peebles2000,Goodman:2000tg,Sahni:1999qe,complexsf8,Arbey:2003sj,Cedeno:2017sou}, for a comprehensive review of this model see also \citeb{rev1,rev2,Marsh:2015xka,rev4,niemeyer2019small,RS}), Dark Energy \citeb{de1,de2,de3} (quintessence \citeb{quintessence,Zlatev:1998tr,Corasaniti:2002vg, de4}, phantom \citeb{de5,de6,de7,de8}, tachyonic SFs \citeb{de9,de10,de11}), inflation \citeb{Linde1982,inf3,inf4,Guth,Lucchin:1984yf,Liddle:1999mq,Ratra:1989uz,DiMarco:2018bnw,inf2}, among others. 

In the inflationary context, it is usually assumed that one or several SFs could have played an important role at early times due to its quantum features \citeb{Linde1982,inf3,inf4,Guth,Lucchin:1984yf,Liddle:1999mq,Ratra:1989uz,DiMarco:2018bnw,inf2} (see also \citep{Vazquez:2018qdg} for a comprehensive review). These fields would be the cause of curvature perturbations that seed gravitational wells for structure in the Universe, as well as gravitational waves. For example, the fluctuations in temperature observed in the CMB are directly attributed to the inflaton quantum fluctuations. Inflation, however, requires a transition to the standard hot big bang cosmology through a process generically dubbed \textit{Reheating} \citeb{reh1,reh2,reh3,reh4,Kofman:1997yn}.

During Reheating, previous to the radiation dominated era and for some kind of inflationary potentials, the SF may start to oscillate around the inflationary potential minimum and behave like a pressure-less fluid. The Reheating epoch, after inflation, represents an important application of Quantum Field Theory (QFT) because it provides a scenario for the origin of the elementary particles of the Standard Model. Then particles interacted with each other up to a state of thermal equilibrium at the so-called \textit{Reheating temperature} $T_r$. 

There are several mechanisms in which the Universe can thermalise through this Reheating scenario. For example, the inflaton condensate could fragment into its own quanta via self-resonance in an oscillating stage and, together with an auxiliary (possibly SM) SF \citeb{Kofman:1997yn,Podolsky:2005bw,GarciaBellido:2002aj,Felder:2001kt,Felder:2000hj,Khlebnikov:1996mc,Greene:1998pb,Dufaux:2006ee}, particles coupled with the inflaton could be resonantly produced off the condensate, leading to prompt thermalisation \citeb{prompt} or a possible oscillon-dominated epoch \citeb{oscillons1,Amin:2011hj,oscillons4,oscillons5}.  This mechanism is called \emph{preheating} (See \citeb{Uzan} for a textbook reference and \citeb{reh3,reh4,Lozanov:2019jxc,Brandenberger1990,Shtanov:1994ce,Kofman_reheating} for  pioneering works). In the absence of resonance, particles are generated by slower perturbative processes, in which perturbations of the condensate laid down during inflation grow linearly with the scale factor after re-entering the horizon \citeb{Starobinsky:1992ts,Tanaka:2007gh, Alcubierre:2015ipa}, giving the possibility of forming compact objects and virialised structures detached from the expanding background (e.g. \citeb{Hogan:1988mp,Kolb:1993hw,Amin:2011hj, Eggemeier:2020zeg}). 

It has been suggested in \citeb{sfdmrh1} (see also \citeb{sfdmrh2,Eggemeier:2020zeg}) that this process of \textit{primordial structure formation} mimics the well-known structure formation process in SFDM models. From the perspective of the dynamical equations used to describe SFDM and the inflaton during a Reheating stage, the two epochs differ only in their initial spectra and parameter values. This analogy suggests the possibility of associating and adapting known results for the SFDM model to the period of Reheating. Of particular interest to this work is the differences between a complex and a real candidate, well characterised in SFDM, that may be of interest in the Reheating scenario. {In the context of SFDM, for example, Ref.~\citeb{complexsf4} studied the exact relativistic cosmological evolution of a complex SF with and without a self-interaction between particles (see also the previous works \citeb{Jetzer,complexsf6,complexsf7,complexsf8} and \citeb{complexsf5} for the generalization for the evolution of this complex SF to an arbitrary potential). In \citeb{complexsf4} the authors showed that the cosmological evolution for the complex SF  differs slightly from its real counterpart, reproducing observations from Big Bang Nucleosynthesis better than the standard  \textit{$\Lambda$-Cold Dark Matter} ($\Lambda$-CDM) model. 

 Moreover, it can be argued that the inflaton SF may have actually been a complex field rather than a real one (see for example \citeb{compinf1,compinf2,compinf3,compinf4,compinf5,compinf6,compinf7,compinf8,compinf9,compinf10,compinf11}). Complex fields appear naturally in extensions of the standard model of particle physics. It is thus natural to look at the effects  complex fields might have in the early Universe. With this in mind, we consider a complex field dominating the energy budget during Reheating. Aiming to explore its characteristics in structure formation, we look at the probability of Primordial Black Hole formation in a complex SF reheating scenario.} Previous works have studied the evolution of real SFs in linear \citeb{Jedamzik,martin_2019,hidalgo} and non-linear stages \citeb{hidalgo_prehe} (see also \citeb{Carr_reheating} for related work). In the present work we aim to extend these studies to the case in which the inflaton field is, instead, a complex SF.

In this paper, we focus on a complex SF dominating the matter density at the end of inflation. We study the evolution of perturbations in the so-called fast oscillating regime, which may take place before or after the interaction with other fields. We show how such scenario entails the existence of an instability regime where perturbations grow without limit. We determine the instability band and address the formation of PBHs under these  conditions. For such purpose, we shall concentrate in the following scenario: PBHs formed during a reheating stage would have evaporated via Hawking radiation if they formed with masses smaller than $10^{-18}~\mathrm{M_\odot}$. Planck mass relics as possible byproducts of the evaporation process would thus be abundant enough to constitute the whole of dark matter. In this work, we use such limit to the abundance of Planck-mass relics of PBHs produced during reheating to bound the period of time that a complex fast oscillating SF can dominate the Universe after the end of inflation. We find that such period cannot extend beyond a few e-folds before PBHs are overproduced, for a wide range of energy levels.  

This paper is organised as follows: In section \ref{Section II} we present the model of a complex SF dominating the energy density of the Universe, right at the end of inflation, and derive the relevant equations and approximate solutions for the background field in a fast oscillating stage. In section \ref{Section III},  we present the perturbative equations at linear level and derive an analytic approximation. In sub-section \ref{Section3.2}, we develop the numerical solutions of our complex SF and compute the relevant power spectra. We estimate in section \ref{Section IV} the formation of PBHs through the spherical collapse model of pressure-less dust by postulating that all collapsed configurations form PBHs in the Press-Schechter count.
In subsection \ref{section4.2}, we constrain the abundance of PBHs formed at the end of inflation by computing their evaporation times and the corresponding Planck relic abundance. Assuming these relics account for all of the dark matter, we derive restrictions to the duration of the fast oscillations period. Finally, we discuss our results and draw conclusions in section \ref{Section V}. For reference, in Appendix \ref{app:A}, we show in detail the derivation of the perturbative treatment for the complex SF. \Correx{Followed by a discussion of the real SF instability in Appendix \ref{app:B}, presented for the sake of comparison.}

\section {\bf The oscillatory field during reheating}\label{Section II}

\begin{figure}[!ht]
\center
\includegraphics[width=0.75\linewidth]{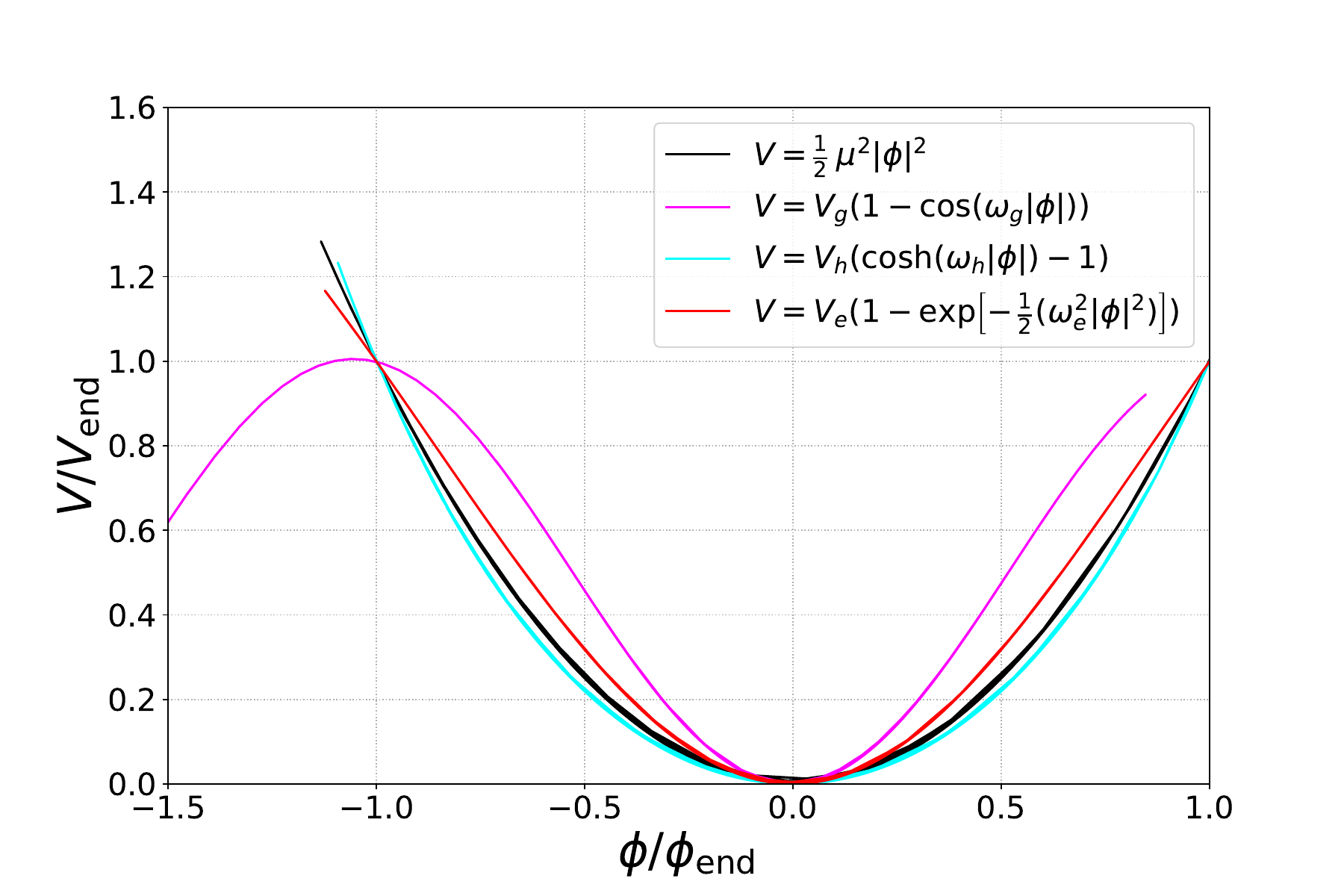}
\caption{Plot of the potential of a few of the models that fit a quadratic behaviour at the minimum. The black curve is the squared potential that we adopt throughout this paper. Both the magenta and cyan lines, representing trigonometric potentials, are typically employed to model SFDM and dark energy from axion-like potentials \citeb{Marsh:2015xka,Urena-Lopez:2015gur,Cedeno:2017sou}, as well as quintessence-like potentials \citeb{Sahni:1999qe,Matos:2000ng,Urena-Lopez:2019xri} and its extensions \citeb{Freese:1990rb, Ross:2009hg,German:2017pfu}. Finally, the red line shows exponential potentials which arise, for example, when families of potentials are derived from the large--$N$ formalism \citeb{Roest:2013fha,Barbosa-Cendejas:2015rba}.}
\label{Fig:Potentials_reheating}
\end{figure}

In this section we present the basic equations necessary to describe a complex SF during the reheating epoch.  We consider the oscillatory phase of the inflaton as the initial stage of the preheating process where, eventually, the energy is transferred to other fields in an explosive particle production. Oscillations take place when the SF lies at a minimum of the potential. Here we model it by a simple harmonic shape 
\begin{equation}
V = \mu^2 |\phi|^2,
\label{potential:phi}
\end{equation}

\noindent where $\mu$ is the mass of the field and also represents the oscillation frequency of the field (as shown below). This potential is associated to a chaotic-like inflationary model, which is ruled-out by current cosmological observations \citeb{akrami2020planck}. However, this form can be taken as the first term in a Taylor expansion near a potential minimum, valid long after the slow-roll stage. The term in Eq.~\eqref{potential:phi} is dominant at the minima of a variety of potentials as those illustrated in Figure \ref{Fig:Potentials_reheating}. {Thus, our model is inspired in a number of inflationary and quintessence models which share the same phenomenology at the oscillatory stage.}

In order to consider both degrees of freedom in the complex field, we treat $\phi$ and its complex conjugate $\phi^*$ as independent fields.
\begin{equation}
(\phi,\phi^*) = \dfrac{1}{\sqrt{2}}\left( \phi_{\rm re} \pm i \phi_{\rm im} \right).
\label{Eq:superposition_scalarfield}
\end{equation} 

The energy-momentum tensor associated to the adopted model,  is given by,
\begin{equation}
T_{\mu \nu} = \partial_\mu \phi^* \partial_\nu \phi + \partial_\mu \phi \partial_\nu \phi^* - g_{\mu \nu}  \underbrace{\left( \partial^\alpha \phi \partial_\alpha \phi^* - \mu^2 \phi \phi^* \right)}_\text{$\mathcal{L}$},
\label{Eq:TSF3}
\end{equation}
where $g_{\mu\nu}$ is the space-time metric. 
This tensor is subject to the conservation equations $\nabla_\nu  T^{\mu \nu} = 0$, and thus the field must satisfy the Klein–Gordon (KG) equation,
\begin{equation}
\Box \phi  + \dfrac{dV}{d \phi^*} = 0.
\label{Eq:KG_complex}
\end{equation}
The energy-momentum tensor is the source term in the Einstein field equations
\begin{equation}
G_{\mu\nu}\equiv R_{\mu\nu}-\frac{1}{2}g_{\mu\nu}R = \kappa T_{\mu\nu},
\label{Eq:Eeq}
\end{equation}
where $\kappa \equiv 8\pi$, $G_{\mu\nu}$ is the Einstein tensor, and $R_{\mu\nu}$ ($R$) is the Ricci tensor (scalar). 

The previous equations constitute the EKG system, which governs the evolution of the complex SF.

\subsection{FLRW background}

Primordial inflation stretches space-time to reach homogeneity and isotropy in a patch of space-time large enough to cover the observable Universe. This justifies modeling the reheating background with the Friedmann-Lemaitre-Robertson-Walker (FLRW) metric, with line element 
\begin{equation}
ds^2 = a(\eta)^2[-d\eta^2 + g^{ij}dx_i dx_j],
\label{Eq:FLRW metric}
\end{equation}

\noindent where {$a(\eta)$ is the scale factor and} the conformal time $\eta$ is related to the cosmic time $t$ by integrating $d\eta = {dt}/{a(t)}$. The conformal Hubble parameter {$\mathcal{H}$} is thus {written in terms of the Hubble parameter $H\equiv d\ln(a)/dt$ as,}
\begin{equation}
\mathcal{H} \equiv \frac{d \ln a}{d \eta} = a H.
\label{Eq:conformal_hubble}
\end{equation}

The relevant Einstein field equations for a flat FLRW universe dominated by a complex SF (i.e. with source tensor in Eq.~\eqref{Eq:TSF3}), are {given from \eqref{Eq:Eeq}}\footnote{The superscript $^{(n)}G$ denotes the n-th order in perturbative expansion.}: 
\begin{equation}
^{\left(0\right)}G^\eta_{\eta} \qquad \rightarrow \qquad \mathcal{H}^2 = \dfrac{8\pi G }{3}\left( \Pi_0 \Pi^*_0  + \mu^2 a^2 \phi_0 \phi^*_0 \right).
\label{Eq:G00_0}
\end{equation}
\begin{equation}
^{\left(0\right)}G^{r}_{r} \qquad \rightarrow \qquad\mathcal{H'} - \mathcal{H}^2 = - 4\pi G\ \Pi_0 \Pi^*_0,
\label{Eq:G11_0}
\end{equation}

\noindent where the background momentum is $\Pi_0 \equiv \phi'_0$ and a prime denotes differentiation with respect to conformal time. Interpreted as a perfect fluid, the energy density and pressure in terms of the homogeneous fields are, respectively,
\begin{equation}
\rho_0 = \dfrac{\Pi_0 \Pi^*_0 }{a^2} + \mu^2 \phi_0 \phi^*_0,
\label{Eq:Density0}
\end{equation}
\begin{equation}
P_0 = \dfrac{\Pi_0 \Pi^*_0 }{a^2} - \mu^2 \phi_0 \phi^*_0.
\label{Eq:Pressure0}
\end{equation}

\noindent The description of SF inhomogeneities through a perfect fluid, however, is incomplete \cite{Christopherson:2012kw}, and we thus follow to the description of the EKG system. The Klein-Gordon equation \eqref{Eq:KG_complex}, at the background level is,
\begin{equation}
\Pi'_0 + 2  \mathcal{H} \Pi_0 + \mu^2 \phi_0 = 0.
\label{Eq:KG_background}
\end{equation}

The KG equation has been studied in detail both for real and the complex SF. Typically, two regimes can be identified which are distinguished by comparing the expansion rate of the Universe and the oscillation frequency of the SF. An inflationary epoch for a real field takes place naturally in the so-called \textit{low oscillating regime}\footnote{It is well-known that a real SF with a sufficiently flat potential undergoes a stiff-matter like stage followed by an inflation era which is an attractor of the EKG equations during the low oscillating regime \citeb{inflatonera1,inflatonera2}. Inflation ends when the slow-roll condition is violated.}, which is identified {through} the condition $\mu\ll H$ (i.e. when the oscillation period of the SF is much larger than the expansion rate of the Universe), whereas for a complex field this inflationary stage is not generic and particular potentials are {required} to obtain an inflationary period in the low oscillating regime\footnote{Ref. \citeb{complexsf5} shows that a complex SF with an attractive self-interaction an inflationary period is expected at early times. However, if a repulsive self-interaction is included, an inflationary epoch is not expected.}. 
    
In the \textit{fast oscillating regime}, when the expansion rate of the Universe is smaller than the oscillation frequency of the SF or ($\mu \gg H$), {both real and complex fields behave effectively as a dust-like component in the Universe.} It is in this regime that the reheating scenario can take place.

Following our motivation to describe our reheating scenario, \Correx{we choose the common expression of the field in terms of a phase instead of the superposition of planes waves}. Hence we \Correx{express} the solution to Eq.~\eqref{Eq:KG_background} as, 
\begin{equation}
\phi_0(t) = C\, a^{-\frac{3}{2}} \exp{\left[ i(\mu t + \psi_0)\right]},
\label{Eq:bg_sol}
\end{equation}

\noindent with $\psi_0$ being a constant, \Correx{real} phase.  \Correx{While the expression in Eq.~\eqref{Eq:bg_sol} is not the most general solution to the complex KG equation, this form is sufficient to satisfy the initial conditions imposed at the end of inflation. It is important to mention that for this particular model, the condition $\mu \gg H$ is not strictly met when $\epsilon = 1$. Nevertheless, we have verified that  the numerical solution rapidly converges towards the approximation \eqref{Eq:bg_sol}.} 
{Since the period of oscillation of the inflaton field follows shortly after the end of the accelerated expansion,} we set initial conditions at $t_{\rm end}$, the time at the end of inflation\footnote{As pointed out in \citeb{sfdmrh2}, quadratic inflation ends when $\phi\sim$ M$_{\rm Pl}$ and $H\sim \mu$, where M$_{\rm Pl}$ is the Planck mass. However, most large field models can be approximated as $V(\phi)\sim \phi^n$, with $n\sim 1$. Considering that inflation ends when $\epsilon\equiv \frac{ {\rm M}_{\rm Pl}^2}{2}\left(\frac{V'}{V}\right)^2\simeq 1$, the endpoint of inflation lies generically at $\phi\sim$ M$_{\rm Pl}$ with $H$ of order $\mu$. Thus, approximating initial conditions as in chaotic inflation may be representative of a larger class of models.}, assuming that the fast oscillations start then, and parametrize the energy scale through $H_{\rm end}$, a free parameter of the model (hereafter the subscript ${\rm end}$ denotes quantities {evaluated} at the end of inflation). 

\begin{figure}[t!]
\centering
\includegraphics[width=0.75\linewidth]{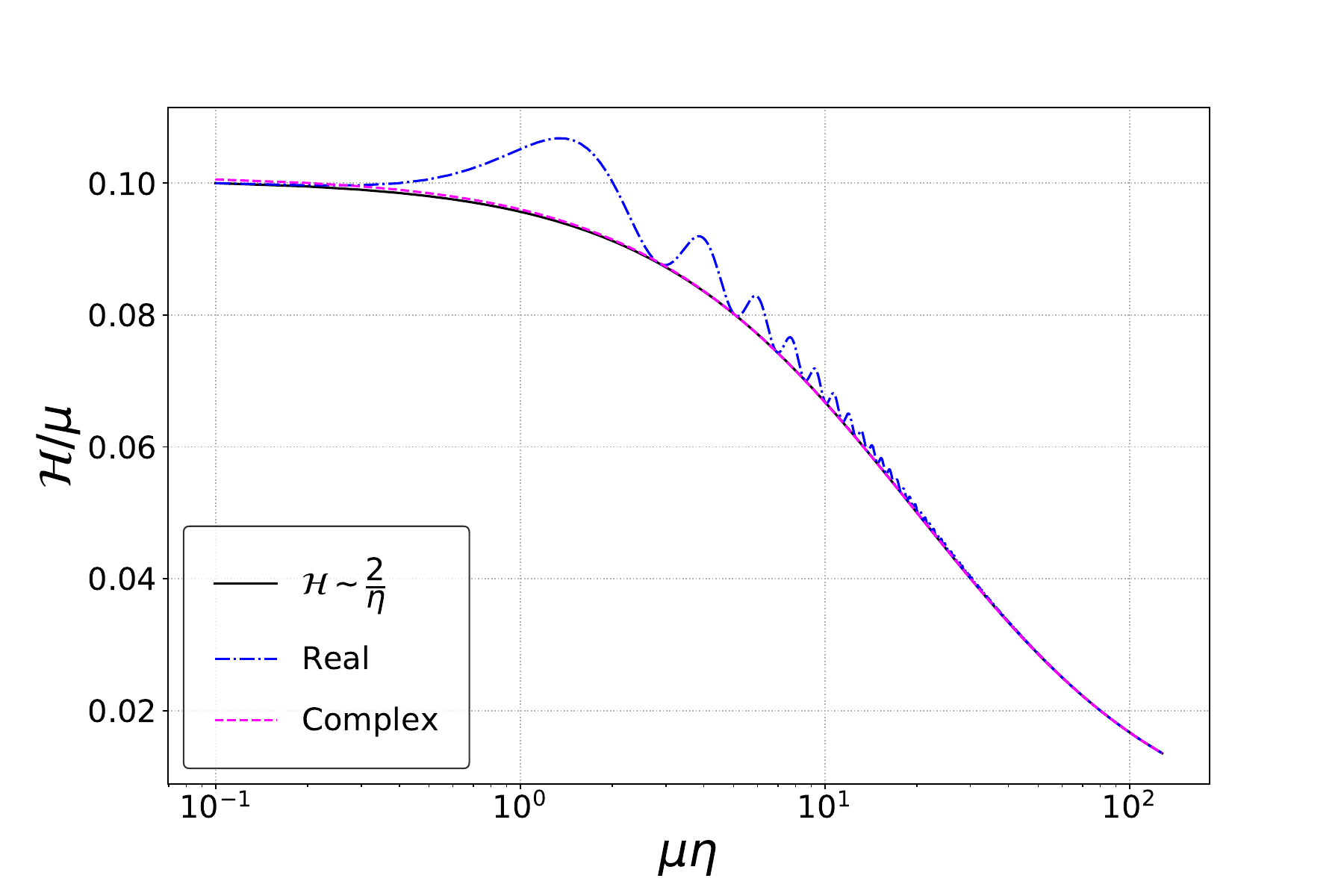}
\caption[Evolution of conformal Hubble parameter]{Evolution of the conformal Hubble parameter $\mathcal{H}/\mu$ as a function of $\mu \eta$ with an initial value of $\mu = 10 H_{\rm end}$ for a real and complex SF oscillating around the minimum of the potential. 
The black line represents the evolution of pressure-less matter, while analytical approximations to the complex and real SF are displayed with a magenta dashed line, and a blue dash-doted line respectively.}
\label{Fig:Hubble_conformal}
\end{figure}

We thus take the field value as,
\begin{equation}
\phi_{\rm end} = \sqrt{\frac{3}{2}} \dfrac{1}{n}\,\text{M}_{\rm Pl},
\label{Eq:phi_end2}
\end{equation}

\noindent with $n$ a large enough number such that $\mu = n H_{\rm end}\gg H_{\rm end}$\footnote{At the end of slow-roll inflation $n= \sqrt{3}/2$
which is of order one. The nominal value $n=10$ used for the initial contidions of our numerical calculations is quickly reached at $N=1.6$ $e$-foldings after the end of inflation. This justifies our identification of the start of oscillations with the end of inflation.}. We can then write, 
\begin{equation}
C =  \sqrt{ \dfrac{{a_{\rm end}^3\left(H^2_{\rm end} - \kappa P_{\rm end}\right)}}{{2\kappa \mu^2}  } }  \exp(-i( \mu t_{\rm end} + \psi_0)).
\label{Eq:C}
\end{equation}

\noindent From \eqref{Eq:Density0} and \eqref{Eq:Pressure0}, a consistent initial value for the momentum is thus,
\begin{equation}
\Pi_{\rm end} \, \Pi^*_{\rm end} = \mu^2 \phi_{\rm end} \, \phi^*_{\rm end}   \quad \rightarrow \quad P_{\rm end}  = 0.
\label{Eq:Pi_background}
\end{equation}

The solution thus mimics a component of non-relativistic particles (pressure-less fluid {or dust}), as illustrated in Figure \ref{Fig:Hubble_conformal}. 

Note that physical quantities are constructed through products of the complex SF and its conjugate, thus cancelling phases and avoiding oscillations. This is clear, for example, in Figure \ref{Fig:Hubble_conformal} where {the evolution of} the Hubble parameter {for the complex SF} shows no oscillations, in contrast with the case of a real SF.

The duration of the oscillatory phase is controlled by a second free parameter of the problem: the number of $e$-foldings elapsed from the end of inflation up to the end of the oscillatory phase, $N_{\rm osc}$. Once this and $H_{\rm end}$ are set, we can determine the energy scale at the end of the fast oscillations phase $\Lambda_{\rm osc}$.  In our case,
\begin{equation}
\Lambda_{\rm osc} \approx \Lambda_{\rm end} \exp \left(-\dfrac{3N_{\rm osc}}{4}\right),
\label{Eq:T_rheating}
\end{equation}

\noindent which, as argued above, is equivalent to pressure-less dust dominating the energy budget throughout the period of fast oscillations. Note that the scale $\Lambda_{\rm osc}$ may be above, or coincide with, the reheating temperature scale $T_r$, which marks the end of the reheating phase\footnote{The reheating temperature $T_r$ is limited to values higher than the Big Bang Nucleosynthesis scale (see e.g. \citeb{Kawasaki:2000en}). The upper limit is $T_r < 10^9 \ {\rm GeV}$ in order to avoid gravitino overproduction \citeb{Kofman,Kawasaki:2000en}. Nevertheless, it is also argued that such temperature is too small to accommodate the standard mechanism of baryogenesis in grand unification theories.}. \Correx{Hereafter we assume the latter possibility unless PBHs dominate the energy budget at $N_{\rm osc}$, in which case it is the PBHs which reheat the universe as we see below}.

Let us finally remark that an important consideration in this work is that the oscillating SF is the inflaton field itself. Implications of our analysis for the hybrid inflation scenario are discussed in the final sections of our paper.

\section {\bf Evolution of Power spectra}\label{Section III}
\subsection{Treatment of perturbations}

We proceed to describe the evolution of the field fluctuations, modelled here as linear perturbations during reheating. {The derivations of the equations presented in this section are presented in appendix~\ref{app:A}.} In our notation, the inhomogeneous fields are split in background quantities (subscript zero) and linear perturbations (subscript one), 
\begin{equation}
\phi = \phi_0(t) + \phi_1(x^i,t).
\label{Eq:PertuSF}
\end{equation}

\noindent {In general}, the perturbative analysis is valid when ${\phi_1}/{\phi_0} \ll 1$. For the geometrical quantities, we take the perturbed comoving FLRW metric in Newtonian gauge, 
\begin{equation}
ds^2 = a(\eta)^2[-(1+2\Phi)d\eta^2 + (1-2\Psi)(dR^2 + R^2 d\Omega^2)].
\label{Eq:Newgaugeconf}
\end{equation}
   
\noindent where $\Phi$ and $\Psi$ are the well-known Bardeen potentials. This gauge, also dubbed the longitudinal gauge, defines hypersurfaces of constant time with no shear (see for reference the reviews \citeb{Malik,Kirklin} and the textbooks \citeb{ellis} and \citeb{lachieze}). The components  $^{\left(1\right)}G^{\eta}_{\eta}$ and $^{\left(1\right)}G^{r}_{\eta}$ of the field equations \eqref{Eq:Eeq} are, respectively,
\begin{equation}
\Delta \Psi - 3\mathcal{H} ( \Psi' + \mathcal{H}\Phi) =  \, 4\pi G( a^2 \mu^2 \phi^*_0 \phi_1 + a^2 \mu^2 \phi_0 \phi^*_1  + \phi'^*_0 \phi'_1 + \phi'_0 \phi'^*_1 - 2\phi'_0 \phi'^*_0 \Phi).
\label{Eq:G00_1}
\end{equation}
\begin{equation}
\Psi' + \mathcal{H} \Phi = 4\pi G(\phi'_0 \phi^*_1 + \phi'^*_0 \phi_1).
\label{Eq:G01_1}
\end{equation}

The above equations are simplified when written in terms of the Mukhanov-Sasaki (M-S) variable. This is defined as,
\begin{equation}
u \equiv a \phi_1 + z\Psi,  
\qquad \text{where} \qquad z \equiv \dfrac{a \phi'_0}{\mathcal{H}}.
\label{Eq:MS}
\end{equation}

\noindent We can thus express the field equations as, 
\begin{equation}
\Delta \left(\dfrac{a^2 \Psi}{\mathcal{H}} \right) 
=  4\pi G \left( u'^* z - u^* z'  + u' z^* - u z'^*  \right).
\label{Eq:G00_manipulate7}
\end{equation}

and,
\begin{equation}
\left( \dfrac{a^2 \Psi}{\mathcal{H}}\right)' = 4 \pi G \left(u^*z  +  uz^* \right).
\label{Eq:G01_manipulate4}
\end{equation}

A common strategy to solve this system is to resort to Fourier space. Then, we can combine the last two equations to derive the evolution of $u_\textbf{k}$ and $u_\textbf{k}^*$,
\begin{equation}
 u''^*_\textbf{k} + \left( k^2  -  \dfrac{z''}{z}\right) u^*_\textbf{k} = 0  \qquad \text{and} \qquad  
 u''_\textbf{k} + \left( k^2 -  \dfrac{z''^*}{z^*} \right)  u_\textbf{k}  = 0 .
\label{Eq:MS_fourier}
\end{equation}

\noindent The above is better-known as the \textit{Mukhanov-Sasaki equation} and each equation is the conjugate of the other. It is important to observe that the scale $|z''/z|^{\frac{1}{2}}$ represents a Jeans{-like} wavenumber $k_J$, meaning that all perturbations with wavelength $2 \pi /k$ above the scale ($\lambda_J = 1/k_J$) are unstable and bound to collapse. \Correx{Note that the analogy with the actual Jeans instability comes exclusively from the similarity with the equations in Fourier space of the matter inhomogeneities in the hydrodynamical representation of the scalar field. A few authors (see for example \citep{niemeyer2019small, complexsf1, Suarez:2015fga}) have argued that the instability found in the M-S variable corresponds to an effective quantum pressure, which is translated to a non-vanishing effective soundspeed, responsible for the dissipation of structures at small scales. While we do not make such correspondence, we note our feature corresponds mathematically to the same kind of instability and hence the name. It is worth mentioning that in the field representation, this effective pressure is well understood and is obtained thanks to the quantum nature of the scalar field.}

In order to {estimate the instability scale and gain insight on} the evolution of $u_\textbf{k}^*$ {(or the corresponding complex conjugate for $u_\textbf{k}$)}, we look at the expression for the instability scale,
\begin{equation}
\dfrac{z''}{z} = 
2\left(\dfrac{\mathcal{H'}}{\mathcal{H}}\right)^2
- \dfrac{\mathcal{H''}}{\mathcal{H}}
- 2\mathcal{H'} 
+ \dfrac{a''}{a}
- \dfrac{2 \mathcal{H'} \phi''_0}{\mathcal{H}\phi'_0}
+ \dfrac{2\mathcal{H}\phi''_0}{\phi'_0}
+ \dfrac{\phi'''_0}{\phi'_0}.
\label{Eq:zbiprime}
\end{equation}
\begin{figure}[h!]
\centering
\includegraphics[width=0.9\linewidth]{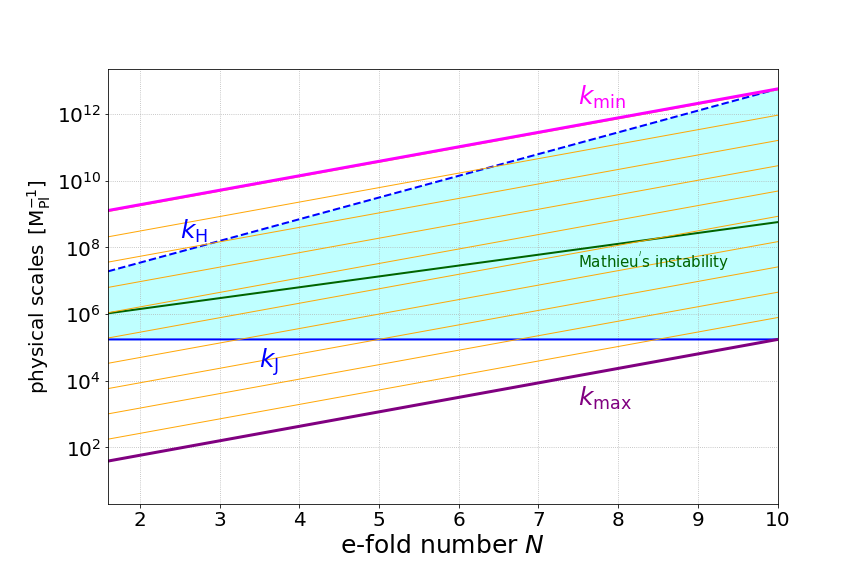}
\caption[Evolution of physical scales during reheating]{The evolution of physical scales is shown as a function of the e-fold number through the fast oscillating regime of a complex SF. The dashed {blue} line is the Hubble horizon. The continuous blue line is the instability scale (Jeans scale). \Correx{The cyan-shaded region between those scales is the instability band}.  \Correx{The orange lines} represent the evolution of $k$-modes.
The {magenta} line $k_{\rm min}$ represents a scale that crosses the Hubble horizon right at the end of the fast oscillations phase. The purple line $k_{\rm max}$ is a scale that never exits the Hubble horizon but reaches the Jeans length only at the last e-fold of reheating. For reference, the green line represents the Mathieu's instability scale for the real SF as presented in Refs.~\citeb{Jedamzik,Alcubierre:2015ipa}.}
\label{Fig:jeans_band_scales}
\end{figure}
\noindent Imposing the fast oscillating regime condition $\mu \gg H$, the above expression reduces to,
\begin{equation}
\dfrac{z''}{z} =  - \mu^2 a^2  + i \, 3 \mathcal{H} \mu a 
\label{Eq:zbiprime_complex2}.
\end{equation}

Note that this expression differs from that of the M-S equation for a real SF under the same approximations, which leads to the well-known Mathieu instability {(see e.g. \citeb{Jetzer,Jedamzik, Alcubierre:2015ipa} for comparison and the discussion of Appendix \ref{app:B})}. This important difference in the instability scale for the real and complex SF is illustrated in Figure~\ref{Fig:jeans_band_scales}, and constitutes one of the main motivations for the present work. \Correx{While in all generality, analytic solutions to the complex KG equation admit periodic solutions for $\mathcal{H}$, and could lead to the Mathieu instability, our analytic approximation to the background field in Eqs.~\eqref{Eq:bg_sol}-\eqref{Eq:C} provides a good estimation of the behaviour right after inflation ends, and yields no oscillating terms in the M-S equation (cf. Eq.~\eqref{Eq:zbiprime_complex2}).}

We proceed to estimate analytic solutions to Eq.~\eqref{Eq:MS_fourier}, approximated for modes well away from the Jeans scale as:
\begin{equation}
 u_\textbf{k} =
  \begin{cases}
  C_1(k) z^* + C_2(k) z^* \displaystyle\int \dfrac{d\eta}{(z^{*})^2}, \qquad & \text{if $\ k^2 \ll \dfrac{z^*{}''}{z^*}$}, \\
  & \\
  \dfrac{\exp{(i k \eta)}}{\sqrt{2k}},  \qquad & \text{if $\ k^2 \gg \dfrac{z^*{}''}{z^*}$}. \\
  \end{cases}
\label{Eq:uk_solution}
\end{equation}

\begin{figure}[!ht]
\centering
\includegraphics[width=0.75\linewidth]{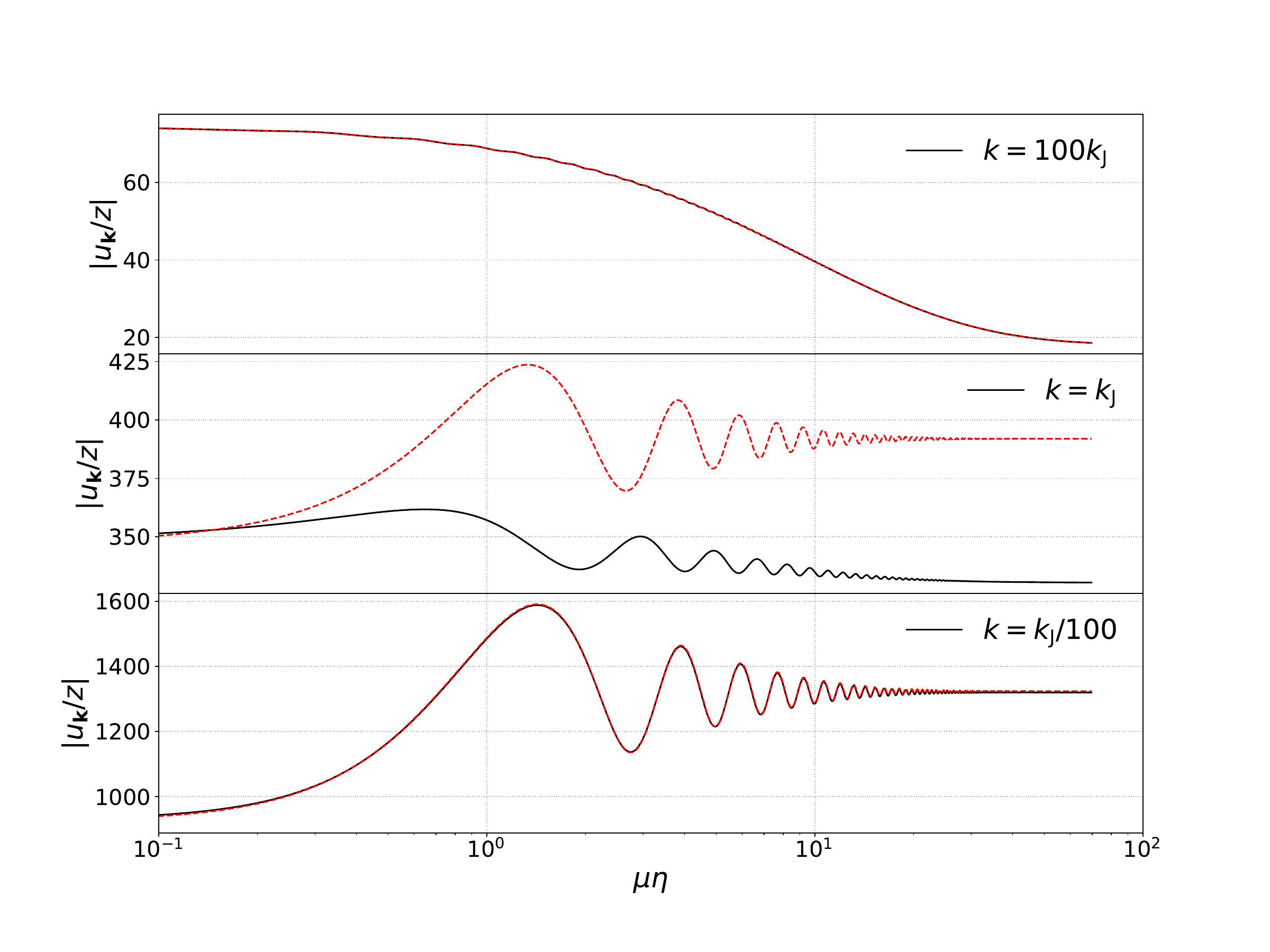}    
\caption[Comparison of perturbations analytical and numerical solution]{{Comparison of the numerical solution to Eq.~\eqref{Eq:MS_fourier} (red dashed line) and the analytic approximation of Eq.~\eqref{Eq:uk_solution} (black solid line). The plot at top represents the stable regime with $ k^2 \gg z^*{}''/z^*$. The bottom plot shows the opposite regime of $ k^2 \ll z^*{}''/z^*$}, where we have employed a soft matching and the initial conditions at the end of inflation. The middle plot just shows the inconsistency of the analytic \eqref{Eq:uk_solution} approximation at the transition (Jeans) scale.}
\label{Fig:u_k}
\end{figure}

Here $C_1$ and $C_2$ are complex integration constants. The behaviour of these solutions is shown in Figure~\ref{Fig:u_k}. Note that the oscillatory solution for large $k$-modes is compatible with the Bunch-Davies vacuum during inflation. In the following, we use this approximate solution to set initial conditions at the end of inflation and evolve the fields numerically during reheating.

\subsection{Setup for numerical evolution}
\label{Section3.2}

Since the difference in expansion of $e$-folds between the end of inflation and the onset of oscillations is of only $N \approx 1.6$, and the evolution of fluctuations is minimal in such transition period, hereafter we assume these two scales coincide and impose initial conditions to the analytic approximation to the M-S variable at the end of inflation, that is, at $k_{\rm end} = a_{\rm end} H_{\rm end}$.
The amplitude of modes at the end of inflation is presented in Ref.~\citeb{bartolo}, where the solutions in terms of the M-S variable after slow-roll inflation are given by,
\begin{equation}
 u_\textbf{k} =
  \begin{cases}
  \dfrac{aH}{\sqrt{2k^3}} \left( \dfrac{k}{aH} \right)^{-\nu_{\phi} +\frac{3}{2}} \exp{(ik \eta)} \qquad & \text{if $\ k^2 \ll \dfrac{z^{*}{''}}{z^*}$}, \\
  & \\
  \dfrac{\exp{(i k \eta)}}{\sqrt{2k}}  \qquad & \text{if $\ k^2 \gg \dfrac{z^{*}{''}}{z^*}$}. \\
  \end{cases}
\label{Eq:uk_solution_inflation}
\end{equation}

Here $\nu_\phi = 2 - n_s/2$ represents a parameter related to the spectral index $n_s$. We have verified the consistency of the approximate solutions in Eq.~\eqref{Eq:uk_solution} against the numerical evolution resulting from taking Eq.~\eqref{Eq:uk_solution_inflation} as initial condition. This is illustrated in Figure \ref{Fig:u_k} for a variety of modes. Indeed, as discussed in Ref.~\citeb{ballesteros}, the decay mode is controlled by the integration constant $C_2(k)$ in Eq.~\eqref{Eq:uk_solution}. Therefore, the amplitude of $C_1(k)$ can be obtained by imposing a smooth matching with the short wavelength solutions or, as in our case, matching the inflation solutions at the end of inflation \eqref{Eq:uk_solution_inflation}.

In principle, we can look at a variety of reheating scenarios by choosing values for the free parameter $H_{\rm end}$, which determines the initial conditions, and control the period of reheating with $N_{\rm osc}$. For a given $k$-mode we have evolved the M-S variable \eqref{Eq:MS} following the M-S equation \eqref{Eq:MS_fourier} and assuming the approximation in Eq.~\eqref{Eq:zbiprime_complex2}. We compute the solution for a range of $k$-modes around the instability scale (Jeans scale) depicted in Figure \ref{Fig:jeans_band_scales}. The results are expressed in terms of the curvature and matter perturbations in the following section.

\subsection{Power spectra}

The curvature perturbation  in the uniform-density gauge $\zeta_\mathbf{k}$ is defined in terms of the M-S variable as,
\begin{equation}
 \zeta_{\,\mathbf{k}} \equiv \dfrac{u_\mathbf{k}}{z}.
   \label{Eq:curvature_zeta}
\end{equation}

From the amplitude of perturbations, we construct the dimensionless power spectrum defined as,
\begin{equation}
 \mathcal{P}_\zeta = \dfrac{k^3}{2\pi} |\zeta_\mathbf{k}|^2.
   \label{Eq:curvature_power_spectrum}
\end{equation} 
 
\begin{figure}[t!]
    \centering
    \begin{subfigure}[b]{0.5\textwidth}
        \includegraphics[width=1.1\textwidth]{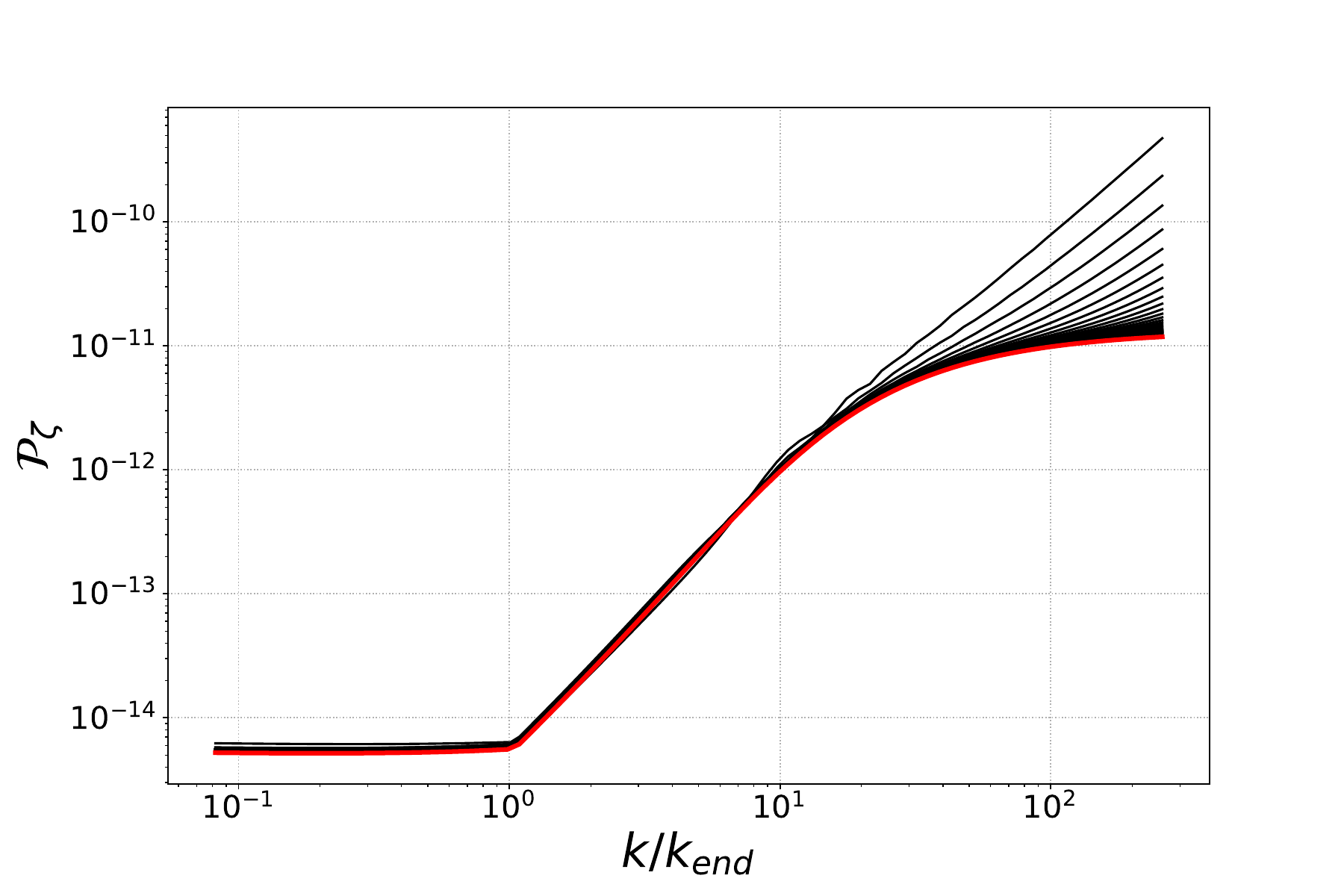}
        \caption{Curvature power spectrum.}
        \label{Fig:curvature_spectrum}
    \end{subfigure}\hfill
    \begin{subfigure}[b]{0.465\textwidth}
        \includegraphics[width=1.1\textwidth]{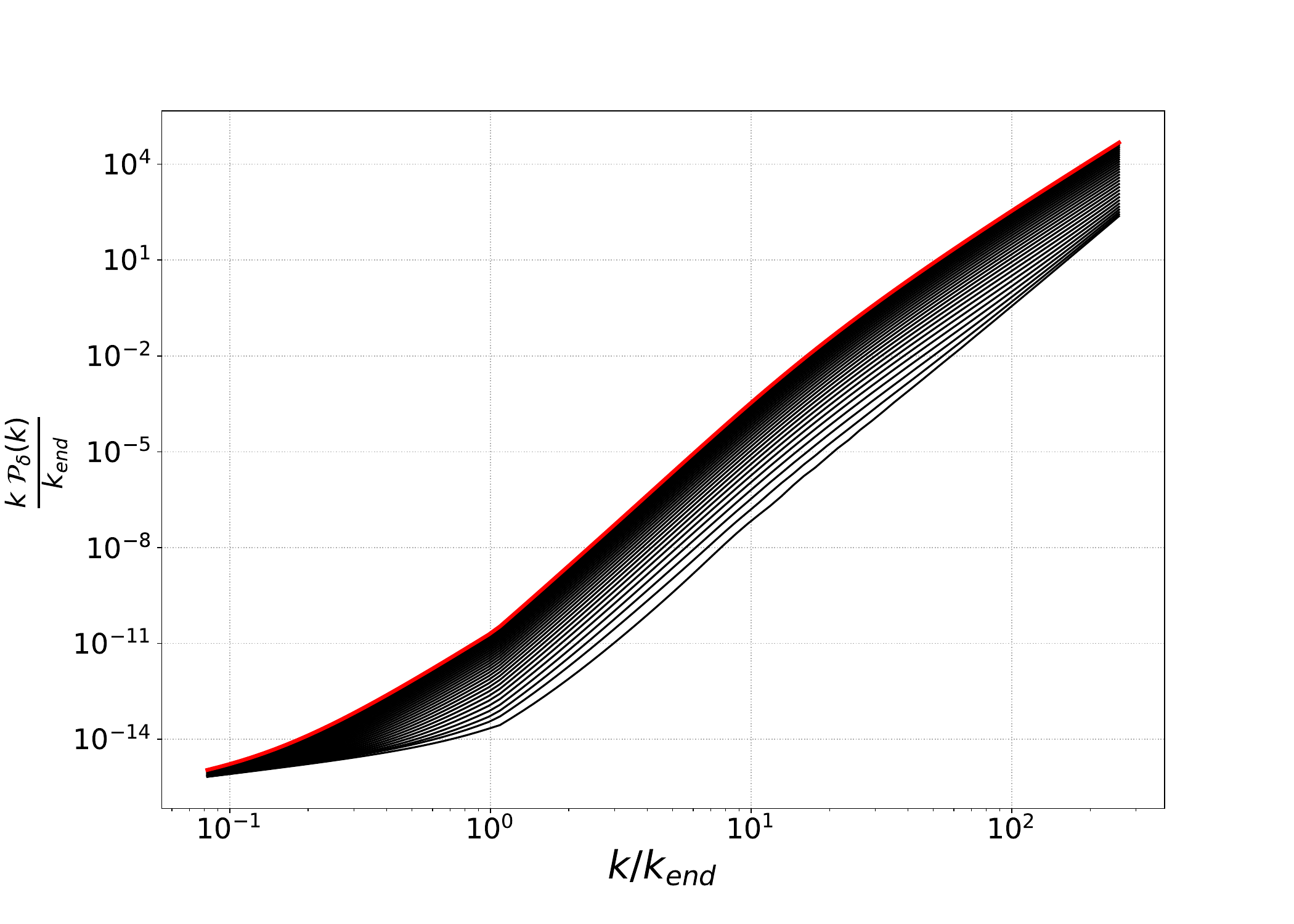}
        \caption{Matter power spectrum.}
        \label{Fig:matter_spectrum}
    \end{subfigure}
    \caption{We compute numerically the evolution of Curvature and Matter Power Spectra through $N = 5$ $e$-folds starting at the end of inflation. In each plot, the sequence of black lines ends with the thick red curve. Note the growth of matter perturbations within the instability band.}
    \label{Fig:power_spectra}
\end{figure}

\noindent The primordial power spectrum has a natural cut-off at the Planck scale in the large-$k$ limit. \Correx{On the other hand, it has been shown in Ref.~\cite{young2014calculating} that modes that remain outside the horizon throughout the PBH formation period are unobservable and should not affect whether a PBH forms or not. For this reason, } \Correx{at the lower end of the spectrum we exclude scales which enter the horizon after the end of the fast oscillation stage, that is, the spectrum is defined up to $k \leq k_{\rm osc} = a_{\rm osc} H_{\rm osc}$. While the evaluation of PBH formation is typically computed in the comoving gauge, where the superhorizon density contrast is suppressed by a factor of order $ k^2 / \mathcal{H}^2$, the cut-off of superhorizon scales in our powerspectrum avoids spurious contributions to this quantity.} We display the numerical evolution of the curvature power spectrum in Figure~\ref{Fig:curvature_spectrum}, illustrating the time-independence of modes larger than the instability scale, i.e.,  $\zeta' \simeq 0$ for $k \ll k_J$ (a characteristic one can immediately infer from the approximate solution in Eq.~\eqref{Eq:uk_solution}, considering only the growing mode). 

The corresponding matter power spectrum is {defined as}, 
\begin{equation}
 \mathcal{P}_\delta = \dfrac{k^3}{2\pi} |\delta_\mathbf{k}|^2.
   \label{Eq:matter_power_spectrum}
\end{equation} 

In a dust{-like} scenario \Correx{and for the $k$-modes within the instability band}, {the matter density perturbation in the longitudinal gauge $\delta_\mathbf{k}$ is related to the M-S variable or the $\zeta$ fluctuations as}:
\begin{equation}
\delta_\mathbf{k}= -\dfrac{2}{5} \left(\dfrac{k^2}{a^2 \, H^2} + 3 \right)\zeta_\mathbf{k}. 
\label{Eq:delta_k}    
\end{equation}

{Taking the curvature perturbation as constant at scales above the Jeans wavelength ($k \ll k_J$) and noting that $a H \propto a^{-1/2}$, then, the relation \eqref{Eq:delta_k} can be approached as}
\begin{equation}
\delta_\mathbf{k} \propto a.
\label{Eq:delta_k_approx}    
\end{equation}

The result of $\delta_\mathbf{k}$ growing with the scale factor $a(t)$, for modes initially inside the Hubble horizon and in the instability band, is also numerically confirmed by \citeb{Jedamzik, auclair2020primordial} and illustrated in Figure \ref{Fig:matter_spectrum}. For such scales, 
\begin{equation}
\dfrac{\mathcal{P}_\delta(t_{\rm osc})}{\mathcal{P}_\delta(t_{\rm end})} = \dfrac{a^2_{\rm osc}}{a^2_{\rm end}} = \exp(2N_{\rm osc}),
\label{Eq:P_delta2}
\end{equation}

where $t_{\rm osc}$ marks the end of the oscillatory phase, after which all of the energy stored in the scalar field is converted to relativistic particles (except for the mass transformed to PBHs which we compute below). 
Now that we have identified the growing mode in the matter field, we can address the formation of PBHs and look at situations where they may be overproduced.

\section{Abundance of Primordial Black Holes }
\label{Section IV}

\subsection{The mass fraction of PBHs}
Since our interest is on the collapse during a reheating period, analogous to a dust-like era, the formation of PBHs is thus not limited by a critical value $\delta_c$ at horizon crossing (see e.g. \citeb{Musco:2012au,Harada_formation}) and a different criterion is required. While the PBH formation in a dust environment is subject to sphericity conditions \citeb{Khlopov:1980mg,Harada:2016mhb}, the criteria required
to track SF inhomogeneities may differ and in this work we assume spherical configurations.
\JCH{Although some threshold values for black hole formation have been put forward in the literature \cite{hidalgo,martin_2019}, in the absence of a definite criterion, we assume that when a spherical inhomogeneity presents a divergent density a black hole is formed. The spherical collapse model marks such moment for a nonlinear dust configuration, through the corresponding linear perturbation critical value $\delta_c = 1.68$.}
\JCH{Thus we adopt this value as an indicator that an overdensity collapses and forms a PBH. Since this occurs during the oscillatory phase of the SF, in our specific exercise, we count configurations of linear overdensities with amplitude above the threshold at $t_{\rm osc}$, discarding any configurations collapsing after this time.}

Let us quickly review how to compute the abundance of PBHs in a given cosmology. The variance $\sigma(R)$ is typically the  size of fluctuations squared, and in terms of the power spectrum we have,
\begin{equation}
\sigma(R)^2 = \int^\infty_0 W^2(kR) \mathcal{P}_\delta(k,t_{\rm end}) \dfrac{dk}{k}.
\label{Eq:sigma_R}
\end{equation}

Here, $W(kR)$ is the Fourier transform of the window
function used to smooth the density contrast. For simplicity, we choose a Gaussian window function, $W(kR) = \exp(-k^2R^2/2)$.   Since the integral in $\sigma(R)^2$ is dominated by the scale $k = 1/R$, we assume without loss of generality, that the primordial power spectrum is scale-invariant. \Correx{Recall that, despite the definition, the integral is limited by a non-vanishing spectrum that runs from $k_{\rm Pl}$ up to $k_{\rm osc}$, and which avoids the spurious contributions of the superhorizon modes.}

We have also verified that for the scales of interest the variance is well approximated by the matter power spectrum. In terms of the horizon mass, \Correx{at the time of horizon crossing},
\begin{equation}
M_H = \dfrac{4\pi}{H_{\rm end}} \left(\dfrac{k_{\rm end}}{k}\right)^3,
\label{Eq:Horizon_mass4}
\end{equation}

\noindent the equivalence is 
\begin{equation}
 \sigma(M,t)^2 \approx \mathcal{P}_\delta(M,t) .
\label{Eq:approx_sigma}
\end{equation}

Obviating the efficiency factor of order one in the precise value of the PBH mass, we assume it to be equal to the horizon mass at cosmological horizon crossing. The initial abundance of PBHs of a given mass $M$ with respect to the critical density is defined as,
\begin{equation}
\beta(M) \equiv \dfrac{\rho_{_{\rm PBH}}(M)}{\rho_c},
\label{Eq:beta_pbh}
\end{equation}

where $\rho_{_{\rm PBH}}$ is the density of PBHs {and $\rho_c$} {is the critical density of the Universe}. \Correx{In order to compute this quantity we recall the Press-Schechter formalism where the fraction of  collapsed objects of mass larger than M is equivalent to the probability that the smoothed density field exceeds the threshold $\delta_c$ \citeb{carr_criterion}.
\begin{equation}
\mathrm{P} [\delta > \delta_c ] = \int^\infty_{\delta_c} \mathrm{P} (\delta) \, d\delta .
\label{Eq:beta_erfc}
\end{equation}}

As argued above,  \Correx{we evaluate this integral at $t_{\rm osc}$ and employ the scale-independent value $\delta_c  = 1.68$ for the linear perturbation, associated to the time of nonlinear collapse. We choose this value as a conservative threshold, given the lack of a definite criterion for the complex SF.}
As in the case of the Scalar Field Dark Matter model, where the power spectrum presents a cutoff at a characteristic scale, one could argue that the threshold amplitude employed in the above formula is scale dependent (see e.g.~\cite{LinaresCedeno:2020dte}). However, $\delta_c$ departs from the dust prescription at scales near the Jeans instability scale, \JCH{which for all realizations of our model lies below the Planck scale, (see Figure~\ref{Fig:Omegas} below).} Thus we can safely take the threshold amplitude for collapse as scale-independent.  

The probability distribution function (PDF)  of the smoothed density field $\mathrm{P} (\delta)$ is assumed to be Gaussian, 
\begin{equation}
\mathrm{P}(\delta) = \dfrac{1}{\sqrt{2\pi} \sigma(M)} 
\exp \left(-\dfrac{\delta^2}{2\sigma(M)^2} \right),
\label{Eq:assumption_gauss}
\end{equation}

\noindent and consequently, \Correx{\begin{equation}
\mathrm{P} [\delta > \delta_c ] = \int^\infty_{\delta_c} \mathrm{P} (\delta) \, d\delta = \dfrac{1}{2} \text{erfc} \left(\dfrac{\delta_c}{\sqrt{2}\ \sigma(M)}\right).
\label{Eq:beta_erfc-bis}
\end{equation}}

\Correx{Since this probability accounts for all configurations of mass at least $M$, we compute the fraction of the total energy density collapsing into PBHs of mass $M$ by differentiating such probability as,
\begin{equation}
\beta(M) = - 2 M \dfrac{\partial R}{\partial M} \dfrac{\partial \mathrm{P} [\delta > \delta_c]}{\partial R},
\label{Eq:beta_renorm}
\end{equation}}

\Correx{where the factor 2 is there to fit estimations of mass fraction from the theory of peaks. Adding the contribution from the complete spectrum of masses of collapsed objects, yields the integrated mass fraction of PBHs:
\begin{equation}
\Omega_{\rm PBH}(t_{\rm osc}) = \int^{\infty}_{0} \beta(M,t_{\rm osc}) \ d\ln M.
\label{Eq:Omega_PBH1}
\end{equation}}

\Correx{The above represents the fraction of energy in the Universe in the form of PBHs at the end of the oscillatory regime. }
\JCH{For the Gaussian distribution, and considering the complete spectrum of masses we have,
\begin{align}
\Omega_{\rm PBH} = \int^{M_{\rm max}}_{M_{\rm min}} \beta(M,t_{\rm osc}) \ d \ln M 
  = &  \ \mathrm{erfc}\left(\frac{\delta_c}{\sqrt2 \ \sigma (M_{\rm min}, t_{\rm osc})}\right) \\    - &    \mathrm{erfc}\left(\frac{\delta_c}{\sqrt2 \ \sigma (M_{\rm max},t_{\rm osc})}\right) \notag.
\label{Eq:Omega_pbh_analytic}
\end{align}}


\JCH{Here $M_{\rm max}$ represents the largest mass which could collapse at the end of the oscillatory regime, and corresponds to the horizon mass at $N_{\rm osc}$ when the scale $k_{\rm min}$ enters the horizon\footnote{\Correx{In practice, however, only the configurations evolved to nonlinear values at $t_{\rm osc}$ are considered to collapse. This prevents inhomogeneities of the horizon size to contribute significantly to the density of PBHs, and no configurations collapsing after $t_{\rm osc}$ are counted either.}}. The lower integration limit is set  by $k_{\rm max}$ through Eq.~\eqref{Eq:Horizon_mass4} or the Planck mass, as discussed below. Note from the definition of the error function that for a finite mass range $\Omega_{\rm PBH} < 1$ and no renormalization scheme is required (the details of renormalization are discussed in Ref.~\cite{auclair2020primordial})}.

\JCH{In our model, the amount  of radiation produced after oscillations is trivially $\Omega_{\rm rad} = 1 - \Omega_{\rm PBH} $. Note that the unbounded growth of inhomogeneities leads to a copious formation of PBHs if the fast oscillation phase lasts long enough (i.e. $N_{\rm osc}$ is large enough). }

The density fraction of PBHs is illustrated in Figure~\ref{Fig:Omegas} for a range of parameter values $(N_{\rm osc}, H_{\rm end})$ employed here. The range of masses of PBHs is such that they are subject to thermal radiation with the \textit{Hawking temperature} given by 
\begin{equation}
T_{\rm PBH} = \dfrac{1}{8 \pi G M} \approx 10^{-7} \left( \dfrac{M}{M_\odot} \right)^{-1} {\rm K}.   
\end{equation}

As a result of this PBHs evaporate in a period of time given by \Correx{their initial mass $M_\star$,}
\Correx{\begin{equation}
t_{\rm eva} =  t_{\rm Pl} \left( \dfrac{M_\star}{M_{\rm Pl}} \right)^3.
\label{eq:time_evaporation}
\end{equation}}

\Correx{If accretion and mergers are ignored, PBHs smaller than $M \sim 10^{15}~\mathrm{g}$ would have evaporated by the present age of the Universe (assuming that the formation time is not a significant fraction of the age of the Universe). Note however, from the masses in Figure \ref{Fig:Omegas}, that PBHs in our model have low masses and evaporate rather quickly. The radiation emitted could alter well-known processes in past stages of the history of our Universe. In the following, we use the bounds to the PBHs relics abundance to constrain the fast oscillations period.}

\subsection{Constraints for reheating}
\label{section4.2}

\noindent Although no confirmed observation of PBHs has been reported, the abundance of PBHs is severely constrained by observations (for compilations see e.g.~\citeb{Josan,Smoot,Carr_2010,carr2020constraints}). The limits to the abundance of PBHs with original mass $M \lesssim 10^{15}\, \rm g$ are derived from effects of the evaporation process, while limits on PBHs of larger mass pertain purely gravitational phenomena.

Since the size of the black hole scales linearly with its mass, Hawking radiation shrinks the BH size as mass is radiated, down to radii close to the Planck scale $\ell_{\rm Pl}$. Physics beyond the Planck scale is uncertain and it is plausible to assume that the evaporation process ceases at the Planck scale, leaving behind a remnant of mass M$_{\rm Pl} = 2.17 \times 10^{-5}$ grams. This is dubbed the \textit{Planck relic}. A population of Planck relics in turn could be abundant enough to constitute a cold dark matter candidate and as such their abundance is constrained by the present dark matter density. Here we look at the possibility that relics could be left over from the PBH production during complex Scalar Field reheating.
\begin{figure}[h!]
    \centering
    \begin{subfigure}[b]{0.49\textwidth}
        \includegraphics[width=1.1\textwidth]{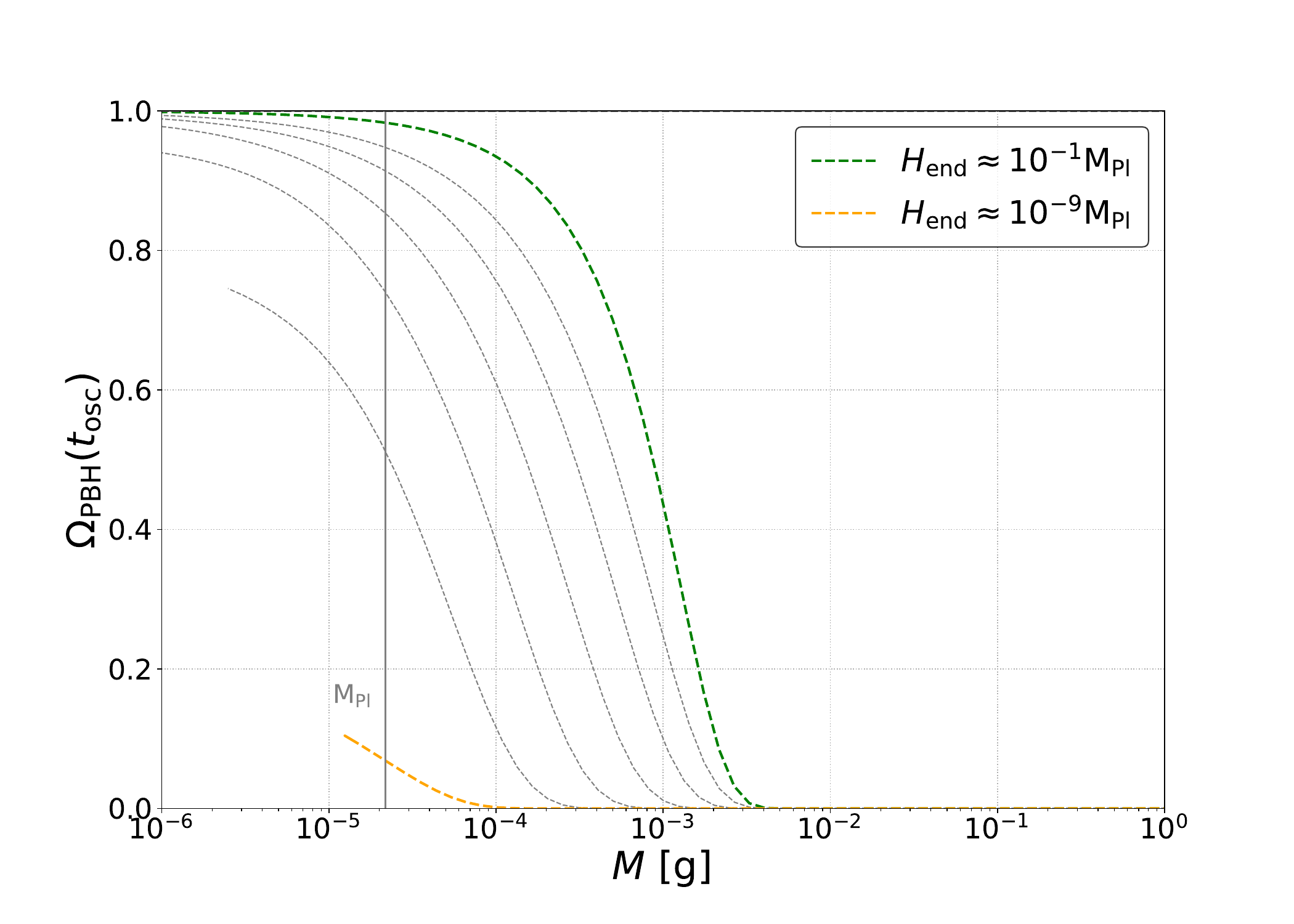}
        \label{Fig:Omega_Hend}
    \end{subfigure}\hfill
    \begin{subfigure}[b]{0.49\textwidth}
        \includegraphics[width=1.1\textwidth]{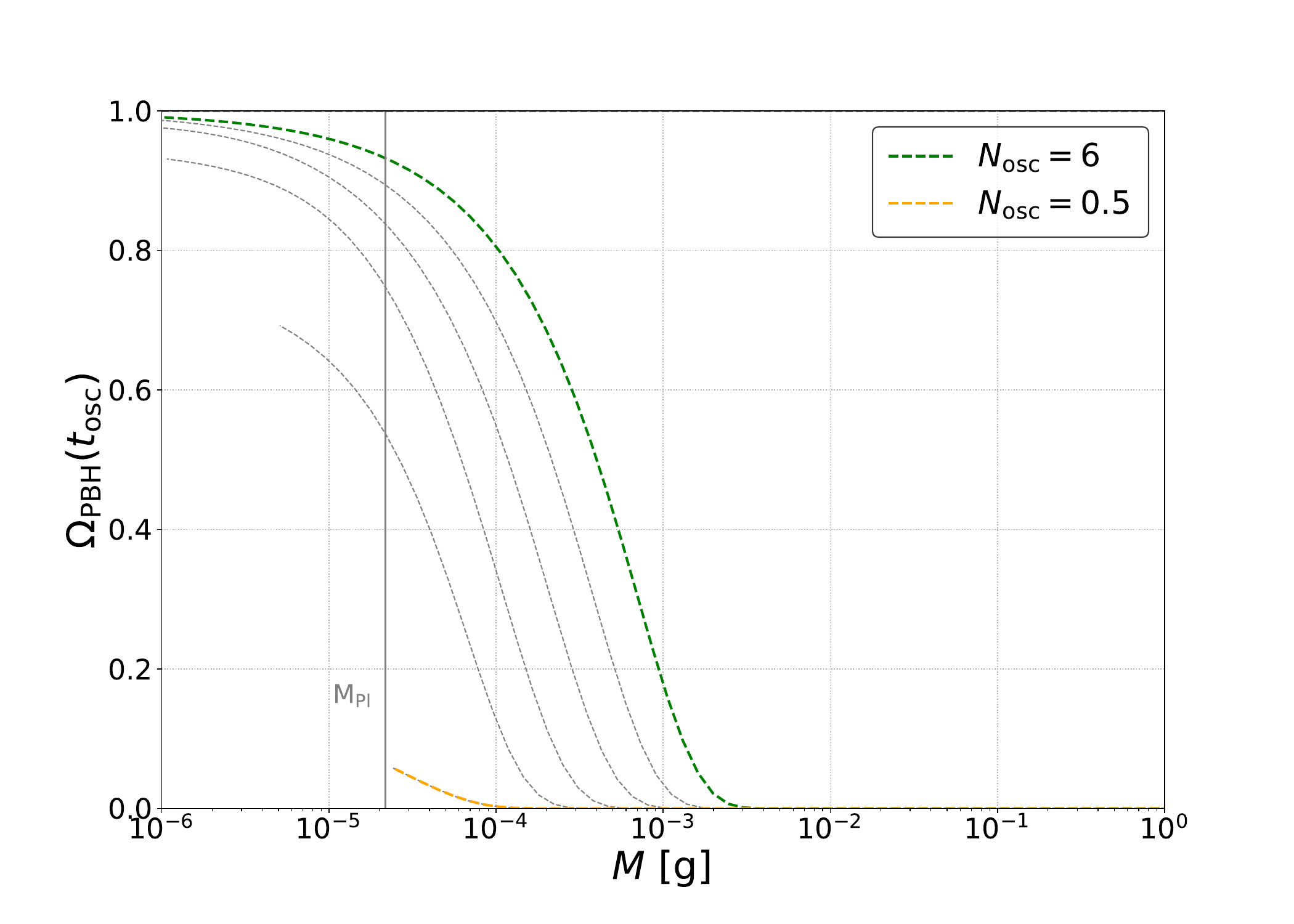}
        \label{Fig:Omega_efolds}
    \end{subfigure}
    \caption{\Correx{Density fraction of PBHs, as a function of PBH mass, varying either of the free parameters in our model: \textit{Left Panel.-} varying the Hubble scale at the end of inflation $H_{\rm end}$, and with $N_{\rm osc} = 5$. \textit{Right Panel.-} Keeping $H_{\rm end} = 10^{-5}~\rm M_{\rm Pl}$  and varying the number of $e$-folds at the end of the fast oscillation stage $N_{\rm osc}$. The cut-off at the Planck mass is shown by the vertical grey line. \JCH{Because of this limit, the lowest values of the mass spectrum do not contribute to $\Omega_{\rm PBH}$. }}}
    \label{Fig:Omegas}
\end{figure}

\Correx{The constraint for Planck relics dark matter appearing before Big Bang nucleosynthesis (BBN) applies to PBHs with initial mass $M < 5.7\times 10^9 $ grams (if they form at an age of the Universe significantly less than 1 sec.). On the other hand, the masses of PBHs studied here range from $\rm M_{\rm Pl}$ to $1~\mathrm{gram}$ as illustrated in Figure~\ref{Fig:Omegas}. Therefore, their evaporation does not spoil BBN processes and they are only bound in abundance to constitute all of the dark matter.}

\Correx{Mathematically, the fractional density of relics integrates the mass fraction of black holes after their evaporation time, reduced by a factor of $\rm M_{\rm Pl}$ for each initial mass $M$, that is, 
\begin{equation}
\Omega_{\rm relic} = \mathfrak{b}(t) \int^{M_{\rm max}}_{M_{\rm min}} \beta(M,t_{\rm osc}) \left( \dfrac{M_{\rm Pl}}{M} \right) \ d \ln M.     
\label{Eq:Omega_relic}
\end{equation}}

\Correx{Here $\mathfrak{b}(t)$ evolves the integrated mass fraction of PBHs in a radiation background as proposed in \cite{auclair2020primordial}. In the range of values sampled here we note that the evolution is almost insignificant  in the interval $[t_{\rm osc}, t_{\rm eva}]$ and we account for it by taking $\mathfrak{b}(t) \sim a(t)$ if $\Omega_{\rm PBH} (t)$ < 1/2, and $\mathfrak{b}(t) = \mathrm{const.}$ otherwise (see also \cite{Carr-Lidsey} for an early account of this evolution in the case of monochromatic production of PBHs).}

\Correx{The constraint for relics to constitute at most all of dark matter is represented by,
\begin{equation}
\dfrac{\Omega_{\rm relic}(t_{\rm eva})}{\Omega_{\rm cdm}(t_{\rm eva})} \leq 1.
\label{Eq:constraint}
\end{equation}}

We consider this Planck relic constraint to impose limits on the free parameters of our reheating model. While all these realizations take $\mu/H_{\rm end} = 10$, the mass could take higher values and the results would remain unchanged.

\begin{figure}[h!]
\centering
\includegraphics[width=\linewidth]{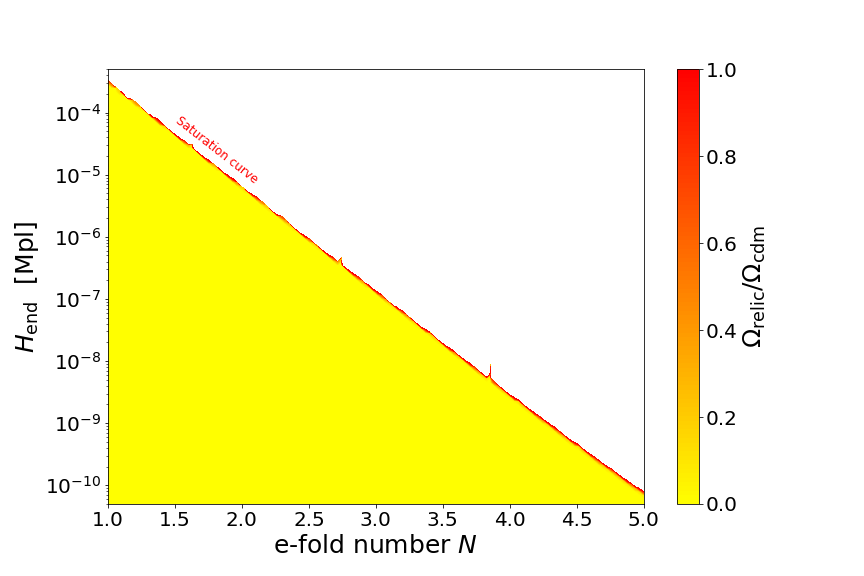}
\caption[Saturation]{\Correx{The exploration of the free parameters of our model is shown in this plot, for realizations with $\mu/H_{\rm end} = 10$. In colors are shown values of $\Omega_{\rm relic}/\Omega_{\rm cdm}$. The \textit{saturation curve} determines a boundary imposed on Planck relics through Eq.~\eqref{Eq:constraint}. Values of the pair $(N_{\rm osc}, H_{\rm end})$ above this line result in an overproduction of PBHs relics}. \Correx{The upper white region of parameter space is completely discarded by this constraint.}}
\label{Fig:saturation}
\end{figure}

We have found a sector of the plane $(N_{\rm osc}, H_{\rm end})$ for which the constraint in \eqref{Eq:constraint} is violated (see Figure \ref{Fig:saturation}).  The direct consequence of this is an inconsistency between the inhomogeneities present during reheating and the dark matter composed of Planck-mass PBHs.

From the results presented in Figure~\ref{Fig:saturation} we infer important consequences for our model. \Correx{If inflation ends at a high-energy scale $H_{\rm end} \geq 10^{-4}\,{\rm M_{\rm Pl}} $  then the fast oscillating regime cannot last for more than one or two $e$-folds. On the other hand, this same phase is admissible for a few $e$-folds when reheating takes place at low scales ($H_{\rm end} \lesssim 10^{-10}\,{\rm M_{\rm Pl}} $, e.g. in low-scale inflation with an auxiliary reheating field \citeb{German:2001tz,Ross:2010fg}).}

\section{Conclusions and outlook}
\label{Section V}

In this paper we have discussed several features on the evolution of inhomogeneities of a complex SF with a harmonic potential in the fast oscillating regime ($\mu \gg H$). We described the evolution of linear perturbations in a reheating scenario and proposed a criterion to account for the formation of PBHs. We finally employed the constraints to the abundance of PBHs through the Planck mass relics remaining after evaporation, and report on limits to the parameters of the model, namely, the scale of the end of inflation $H_{\rm end}$ and the duration of the fast oscillation phase $N_{\rm osc}$.

Differences with the reference case of a real SF appear from the background level.  \Correx{ We have shown that the conditions at the end of inflation allow for solutions that present no oscillation in the background matter density, in contrast with the case of the real SF (and as illustrated in Figure~\ref{Fig:Hubble_conformal}). This important difference with the real case implies an instability scale which differs in size and nature from that of the real field (the instability scale associated to the Mathieu equation presented in Appendix~\ref{app:B}). Our case shows a wider range of $k$-modes covered by the instability band. }
Consequently, complex SF inhomogeneities can collapse with smaller sizes than in the real SF reference case.

After evolving numerically linear fluctuations of the complex SF we propose the critical value of the spherical collapse model ($\delta_c = 1.68$) as the threshold amplitude for which inhomogeneities form Primordial Black Holes. The PBHs in our scenario are formed with masses relevant to the Planck relics constraint in Eq.~\eqref{Eq:constraint}; a bound which imposes limits to the parameter space of our model ($N_{\rm osc}$, $H_{\rm end}$). This final result is illustrated in Figure \ref{Fig:saturation}, presented in the previous section.

The results of this work show a path to constrain models of reheating at stages equivalent to the fast oscillating complex SF. 
While the field here considered has been directly linked to the inflaton itself, such condition can be relaxed and the results are still valid for any SF in an oscillating stage previous to the hot Big Bang. Of particular interest is the class of hybrid inflation potentials for which the field responsible of ending inflation is independent of the inflaton and our setup could be suitable for such scenarios (see e.g. \citeb{Linde:1993cn,Ross:2010fg}). A whole sector of this class could be ruled out in the cases where PBHs are overproduced.

\newpage 

Extensions and improvements to this work are in sight, some of which are:

\begin{itemize}

    \item[$\star$] 
    Considering more involved models of reheating including, e.g. self-interact\-ing potentials (see \citeb{reh4}). This would lead to a new set of scenarios to be constrained by the abundance of PBHs produced during such period. Another, more realistic possibility, is to consider more than one SF, where PBHs also seem to form copiously \citeb{hidalgo_prehe}. 

    \item[$\star$] 
    Assuming the complete inflation model (e.g. \citeb{mishra}), it is possible to constrain the primordial power spectrum with the bounds to the PBH abundance reported here. Changing the prescriptions to the primordial power spectrum at small scales by including the positive running of the spectral index, may lead to more severe constraints at the reheating stages here studied. 
    
    \item[$\star$] 
    With the solutions obtained at first order, it is possible to set semi-analytic initial conditions to the non-linear gravitational collapse, and subsequent black hole formation. It remains to look at the full non-linear collapse and establish a critical value for the collapse of the cosmological SF. This is likely to be a model-dependent value with particular consequences for PBHs production; a property that emphasises the importance of PBHs as a tool for testing conditions of the early Universe.  
\end{itemize}

\acknowledgments
\Correx{We want to thank the anonymous referee for several insightful comments that improved our paper.} The authors acknowledge sponsorship from CONACyT through grant CB-2016-282569. KC acknowledges financial support from master's studentship provided by CONACyT and PAPIIT-UNAM, IA102219 (\textit{Cosmología observacional de materia y energía oscura}). JCH acknowledges support from UNAM PAPIIT grant IN-107521 (\textit{Sector oscuro y Agujeros Negros Primordiales}) and from FORDECYT-PRONACES project No. 304001/2020.

\newpage 
\appendix

\section{The Mukhanov-Sasaki equation for a Complex scalar field}
\label{app:A}

In this appendix we derive the dynamical equation of the perturbations in detail. We start from equations \eqref{Eq:G00_1} and \eqref{Eq:G01_1}:  
\begin{equation}
\Delta \Psi - 3\mathcal{H} ( \Psi' + \mathcal{H}\Phi) =   4\pi G( a^2 \mu^2 \phi^*_0 \phi_1 + a^2 \mu^2 \phi_0 \phi^*_1  + \phi'^*_0 \phi'_1 + \phi'_0 \phi'^*_1 - 2\phi'_0 \phi'^*_0 \Phi ).
\label{Eq:A_G00_1}
\end{equation}
\begin{equation}
\Psi' + \mathcal{H} \Phi = 4\pi G(\phi'_0 \phi^*_1 + \phi'^*_0 \phi_1).
\label{Eq:A_G01_1}
\end{equation}

Additionally, due to azimuthal symmetry, $^{\left(1\right)}G^{\theta}_{\theta}$ and $^{\left(1\right)}G^{\varphi}_{\varphi}$ are the same. From this, we can show that there is no anisotropic stress tensor $\Pi_{ij}$ at 1st order. The proof comes from taking the differences between, 
\begin{equation}
G^{r}_{r} - G^{\theta}_{\theta} = 8\pi G (\, T^{r}_{r} - \, T^{\theta}_{\theta}\, ).
\label{Eq:A_differences_1st}
\end{equation}

This guarantees that,
\begin{equation}
 \dfrac{d^2\left(\Psi - \Phi  \right)}{dR^2} - \dfrac{1}{R}\dfrac{d \left(\Psi - \Phi  \right)}{dR}  =  \left(\dfrac{28\pi G}{1-2\Phi}\right)\dfrac{d\phi_1 }{dR} \dfrac{d\phi^*_1}{dR}.
\label{Eq:A_no_anisotropic}
\end{equation}

The right hand side is a second order term, so we can safely neglected it at linear order and this implies,
\begin{equation}
   \Phi = \Psi.\, \iff \, \Pi_{ij} = 0.
    \label{Eq:A_no_anisotropic2}
\end{equation}

Thus, Eq.~\eqref{Eq:A_G00_1} can be cast as, 
\begin{align}
\Delta \left(\dfrac{a^2 \Psi}{\mathcal{H}} \right) = & \, \dfrac{4\pi G a^2}{\mathcal{H}}\left( a^2 \mu^2 \phi^*_0 \phi_1 + a^2 \mu^2 \phi_0 \phi^*_1  + \phi'^*_0 \phi'_1 + \phi'_0 \phi'^*_1 \right. \notag \\ &  \left. - 2\phi'_0 \phi'^*_0 \Psi \right) 
+ 3 a^2 ( \Psi' + \mathcal{H}\Psi). 
\label{Eq:A_G00_manipulate0}
\end{align}

From now on, the background equations will be useful in order to reduce our perturbative equations. For instance, we can substitute \eqref{Eq:G00_0} in the above equation and find that,
\begin{align}
\Delta \left(\dfrac{a^2 \Psi}{\mathcal{H}} \right) = & \, \dfrac{4\pi G a^2}{\mathcal{H}}\left( a^2 \mu^2 \phi^*_0 \phi_1 + a^2 \mu^2 \phi_0 \phi^*_1  + \phi'^*_0 \phi'_1 + \phi'_0 \phi'^*_1  \right) \notag \\ & + 3 a^2  \Psi' + \dfrac{8\pi G a^4 \mu^2}{\mathcal{H}} \phi_0 \phi^*_0 \Psi.
\label{Eq:A_G00_manipulate1}
\end{align}

We rearange the term $3 a^2  \Psi'$ by multiplying a factor one, derived from the background Eq.~\eqref{Eq:G00_0},
\begin{align}
\Delta \left(\dfrac{a^2 \Psi}{\mathcal{H}} \right) = & \, \dfrac{4\pi G a^2}{\mathcal{H}}\left( a^2 \mu^2 \phi^*_0 \phi_1 + a^2 \mu^2 \phi_0 \phi^*_1  + \phi'^*_0 \phi'_1 + \phi'_0 \phi'^*_1 \right)  \notag \\ & + \dfrac{8\pi G a^4 \mu^2}{\mathcal{H}} \phi_0 \phi^*_0 \Psi + 3 a^2 \Psi'  \left[ \dfrac{8\pi G \left( \phi'_0 \phi'^*_0 + a^2 \mu^2 \phi_0 \phi^*_0\right)}{3 \mathcal{H}^2} \right]. 
\label{Eq:A_G00_manipulate2}
\end{align}

Then,
\begin{align}
\Delta \left(\dfrac{a^2 \Psi}{\mathcal{H}} \right) =
&  \, 4\pi G \left[  \dfrac{a^4 \mu^2}{\mathcal{H}} \left(  \phi^*_0 \phi_1 + \phi_0 \phi^*_1 \right) 
+ \dfrac{a^2}{\mathcal{H}}\left( \phi'^*_0 \phi'_1 + \phi'_0 \phi'^*_1 \right) 
  + \dfrac{ 2 a^2 \phi'_0 \phi'^*_0 \Psi'}{\mathcal{H}^2}  \right]
  \notag \\ & 
  + \dfrac{8\pi G a^4 \mu^2}{\mathcal{H}^2} \phi_0 \phi^*_0 \Psi'  + \dfrac{8\pi G a^4 \mu^2}{\mathcal{H}} \phi_0 \phi^*_0 \Psi . 
\label{Eq:A_G00_manipulate3}
\end{align}

In the last two terms, outside of square parenthesis, we substitute Eq.~\eqref{Eq:A_G01_1},  
\begin{align}
\Delta \left(\dfrac{a^2 \Psi}{\mathcal{H}} \right) = 
&  \ 4\pi G \left[  \dfrac{a^4 \mu^2}{\mathcal{H}} \left(  \phi^*_0 \phi_1 + \phi_0 \phi^*_1 \right) 
+ \dfrac{a^2}{\mathcal{H}}\left( \phi'^*_0 \phi'_1 + \phi'_0 \phi'^*_1 \right)\right. 
  \notag \\ &  + 2 a^2 \phi'_0 \phi^*_1 - 2 a^2 \phi'_0 \phi^*_1  
   + 2 a^2 \phi'{}^*_0 \phi_1 - 2 a^2 \phi'{}^*_0 \phi_1 \notag \\ & \left. + \dfrac{ 2 a^2 \phi'_0 \phi'{}^*_0 \Psi'}{\mathcal{H}^2}  \right]   + \dfrac{8\pi G a^4 \mu^2}{\mathcal{H}^2} \phi_0 \phi^*_0 \left[ 4\pi G(\phi'_0 \phi^*_1 + \phi'{}^*_0 \phi_1) \right].
\label{Eq:A_G00_manipulate4}
\end{align}

We can complete the KG equation, so factorising we obtain, 
\begin{align}
\Delta \left(\dfrac{a^2 \Psi}{\mathcal{H}} \right) = & \ 4\pi G \left[\dfrac{a^2 \phi'_0 \phi'^*_1}{\mathcal{H}}  + \dfrac{a^2 \phi'^*_0 \phi'_1}{\mathcal{H}} + \dfrac{2 a^2 \phi'_0 \phi'^*_0 \Psi'}{\mathcal{H}^2} \right.  \notag \\ & + \dfrac{8\pi G a^4 \mu^2}{\mathcal{H}^2} \phi_0 \phi^*_0 \left[\cancel{\phi'_0 \phi^*_1} + \bcancel{\phi'^*_0 \phi_1} \right]
 - \dfrac{a^2 \phi^*_1}{\mathcal{H}} \left( - 2\mathcal{H}\phi'_0  - a^2 \mu^2 \phi_0 \right) \notag \\ &   + a^2 \phi'_0 \phi^*_1 \left[1  - \dfrac{8\pi G \left( \phi'_0 \phi'^*_0 + \cancel{a^2 \mu^2 \phi_0 \phi^*_0}\right)}{\mathcal{H}^2} \right]
 - \dfrac{a^2 \phi_1}{\mathcal{H}} \left(- 2\mathcal{H}\phi'^*_0 \right. \notag \\ & \left. \left.  - a^2 \mu^2 \phi^*_0 \right) + a^2 \phi'^*_0 \phi_1 \left[1 - \dfrac{8\pi G \left( \phi'_0 \phi'^*_0 + \bcancel{a^2 \mu^2 \phi_0 \phi^*_0}\right)}{\mathcal{H}^2} \right] \right].  
\label{Eq:A_G00_manipulate5}
\end{align}

Moreover, we can substitute from Eq.~\eqref{Eq:G11_0} the expression,
\begin{equation}
\dfrac{\mathcal{H}'}{\mathcal{H}^2} - 1 = -\dfrac{8\pi G}{\mathcal{H}^2}  \phi'_0 \phi'^*_0.
\label{Eq:A_G11_0_rw}
\end{equation}

Then rearranging,
\begin{align}
\Delta \left(\dfrac{a^2 \Psi}{\mathcal{H}} \right) = & \ 4\pi G \left[\dfrac{a^2 \phi'_0 \phi'^*_1}{\mathcal{H}} + \dfrac{a^2 \phi'_0 \phi'^*_0 \Psi'}{\mathcal{H}^2} - \dfrac{a^2 \phi^*_1 \phi''_0}{\mathcal{H}} + a^2 \phi'_0 \phi^*_1 \dfrac{\mathcal{H'}}{\mathcal{H}} \right. \notag \\ &    + \dfrac{a'a \phi'_0 \phi^*_1}{\mathcal{H}} - \dfrac{a'a \phi'_0 \phi^*_1}{\mathcal{H}}  + \dfrac{a^2 \phi'^*_0 \phi'_1}{\mathcal{H}} + \dfrac{ a^2 \phi'_0 \phi'^*_0 \Psi'}{\mathcal{H}^2} \notag \\ & \left. - \dfrac{a^2 \phi_1 \phi''^*_0}{\mathcal{H}} + a^2 \phi'^*_0 \phi_1 \dfrac{\mathcal{H'}}{\mathcal{H}}  + \dfrac{a'a \phi'^*_0 \phi_1}{\mathcal{H}} - \dfrac{a'a \phi'^*_0 \phi_1}{\mathcal{H}}  \right].
\label{Eq:A_G00_manipulate6}
\end{align}

The above equation is enormously simplified when written in terms of the M-S variable Eq.~\eqref{Eq:MS}. In that case, Eq.~\eqref{Eq:A_G00_manipulate5} takes the form,
\begin{equation}
\begin{aligned}
\Delta \left(\dfrac{a^2 \Psi}{\mathcal{H}} \right) ={}
&  4\pi G \left[  a'z \phi^*_1 + az\phi'^*_1 + z \Psi' z^* - a \phi^*_1 z'   +z \Psi z'^* - z\Psi z'^*  \right.\\
& \left. + a' z^* \phi_1 + a z^* \phi'_1 + z \Psi' z^* - a \phi_1 z'^*   +z^* \Psi z' - z^*\Psi z'  \right].
\end{aligned}
\label{Eq:A_G00_manipulate6_2}
\end{equation}
We can thus reduce the equation considerably by employing the M-S variable, this may be expressed as, 
\begin{equation}
\Delta \left(\dfrac{a^2 \Psi}{\mathcal{H}} \right) 
=  4\pi G \left( u'^* z - u^* z'  + u' z^* - u z'^*  \right).
\label{Eq:A_G00_manipulate7}
\end{equation}

This first result is one of two elements required to find the evolution equation for the M-S variable. Proceeding in the same fashion with Eq.~\eqref{Eq:A_G01_1},
\begin{equation}
\dfrac{a^2 \Psi'}{\mathcal{H}} + a^2 \Psi  = \dfrac{4\pi G a^2}{\mathcal{H}} (\phi'_0 \phi^*_1 + \phi'^*_0 \phi_1).
\label{Eq:A_G01_manipulate0}
\end{equation}
\begin{align}
\dfrac{a^2 \Psi'}{\mathcal{H}} + 2a^2 \Psi - a^2 \Psi 
& + \dfrac{8\pi G a^2}{\mathcal{H}^2} \phi'_0 \phi'^*_0 \Psi 
 = \notag \\ & + \dfrac{4\pi G a^2}{\mathcal{H}} (\phi'_0 \phi^*_1 + \phi'^*_0 \phi_1) 
+ \dfrac{8\pi G a^2}{\mathcal{H}^2} \phi'_0 \phi'^*_0 \Psi.
\label{Eq:A_G01_manipulate1}
\end{align}

We can now use \eqref{Eq:A_G11_0_rw} in the above to obtain,
\begin{equation}
\dfrac{a^2 \Psi'}{\mathcal{H}} + 2a^2 \Psi - \dfrac{a^2\mathcal{H}'}{\mathcal{H}^2} \Psi 
 = \dfrac{4\pi G a^2}{\mathcal{H}} \left( \phi'_0 \phi^*_1 + \phi'^*_0 \phi_1 
+ \dfrac{2 \phi'_0 \phi'^*_0 \Psi }{\mathcal{H}} \right).
\label{Eq:A_G01_manipulate2}
\end{equation}

The left hand side is a total derivative and reduces to,
\begin{equation}
\left( \dfrac{a^2 \Psi}{\mathcal{H}}\right)' = 4\pi G \left( a z \phi^*_1 + a z^* \phi_1
+ 2 z z^* \Psi  \right).
\label{Eq:A_G01_manipulate3}
\end{equation}

In terms of the M-S variable  \eqref{Eq:MS}, we have,
\begin{equation}\left( \dfrac{a^2 \Psi}{\mathcal{H}}\right)' = 4 \pi G \left(u^*z  +  uz^* \right).
\label{Eq:A_G01_manipulate4}
\end{equation}

This formulation is particularly convenient to relate the conformal time derivative of \eqref{Eq:A_G00_manipulate7} and the Laplacian of \eqref{Eq:A_G01_manipulate4},
\begin{equation}
\left( u'^* z - u^* z' \right)' + \left(u' z^* - u z'^*  \right)' - \Delta \left(u^*z  +  uz^* \right) = 0.
\label{Eq:A_MS_eq1}
\end{equation}

Reducing,
\begin{equation}
 u''^* z - u^* z''  + u'' z^* - u z''^*   - z \Delta u^*  -  z^* \Delta u = 0,
\label{Eq:A_MS_eq2}
\end{equation}

and with a suitable factorization we have, 
\begin{equation}
\left[ u''^* - \Delta u^* -  \dfrac{z''}{z} u^* \right] z +  \left[ u'' - \Delta u - \dfrac{z''^*}{z^*} u  \right] z^* = 0. 
\label{Eq:A_MS_eq3}
\end{equation}

{In order for the} above equation to be fulfilled, the following {differential equations} must be satisfied {independently},
\begin{equation}
 u''^* - \Delta u^* -  \dfrac{z''}{z} u^* = 0,  \qquad \text{and} \qquad  
 u'' - \Delta u - \dfrac{z''^*}{z^*} u  = 0.
\label{Eq:A_MS_eq4}
\end{equation}

As we shall show in the following appendix, while in the complex case the equations are slightly more complicated than for the real case, their structure is similar. These and other demonstrations are discussed in more detail in Ref.~\citeb{Karim_thesis}.

\section{The Mathieu instability of a Real scalar field}
\label{app:B}

\noindent In this Appendix we examine and discuss the real SF during the oscillatory phase. The evolution of real SF fluctuations during reheating has been studied by several authors (see for example  \citeb{Alcubierre:2015ipa}, \citeb{Jedamzik}, \citeb{martin_2019} and \citeb{hidalgo}). For the background, the solution of a real SF is commonly presented as,
\begin{equation}
\phi_0(t) = A \, a^{-\frac{3}{2}} \sin{(\mu t + \theta_0)}, 
\label{Eq:bg_sol_real}
\end{equation}

with the integration constant $A$. As illustrated in  Figure \ref{Fig:Hubble_conformal}, when we average this solution over a single period of oscillation $1/\mu$, we can write
\begin{equation}
\langle \rho \rangle = \dfrac{1}{2} \langle \Pi^2 \rangle + \dfrac{1}{2} \mu^2 \langle \phi^2 \rangle \approx \mu^2 \langle \phi^2 \rangle \approx \dfrac{\mu^2 \phi^2_0}{2} a^{-3} + \mathcal{O}^2\left(\dfrac{H}{\mu}\right).
 \label{Eq:avg_density}
 \end{equation}
\begin{equation}
 \langle P \rangle = \dfrac{1}{2} \langle \Pi^2 \rangle - \dfrac{1}{2} \mu^2 \langle \phi^2 \rangle \approx \dfrac{9\phi^2_0 H^2}{16a^3} \approx 0.
 \label{Eq:avg_pressure}
 \end{equation}

If the oscillating real SF dominates the Universe for sufficiently long time, it will effectively behave as pressure-less fluid. Consequently, through of the perturbative regime, and employing the M-S variable \eqref{Eq:MS} and the same methods of Appendix~\ref{app:A}, we arrive at the M-S equation, 
\begin{equation}
 u''_\textbf{k} + \left( k^2  -  \dfrac{z''}{z}\right) u_\textbf{k} = 0.
 \label{Eq:jeans_real_eq}
 \end{equation}

Notice from Figure \ref{Fig:Hubble_conformal} that a real SF follows closely the evolution of the standard CDM universe.  Thus, expression  for the \textit{Jeans\Correx{-like} wavenumber}, Eq.~\eqref{Eq:zbiprime}, can be simplified by making use of background equations and derive an explicit instability scale in the evolution. According to \citeb{martin_inflation,martin_2019,Alcubierre:2015ipa,hidalgo}, Eq.~\eqref{Eq:zbiprime} can be approached as,
\begin{equation}
\dfrac{z''}{z} \approx - \mu^2 a^2 \left[ 1 {+} 6\left( \dfrac{H}{\mu} \right) \sin(2 \mu t) + \mathcal{O}^2\left(\dfrac{H}{\mu}\right) \right].
\label{Eq:zbiprime_real}
\end{equation}

Introducing a new change of variable $x \equiv \mu t + \pi/4$ on \ref{Eq:jeans_real_eq}  \Correx{and defining the new variable $\tilde u_\text{k} \equiv a^{1/2}u_\textit{k}$} one obtains the Mathieu equation as follows, 

\Correx{\begin{equation}
\dfrac{d^2 \tilde u_\textbf{k}}{dx^2} + \left[ A(\textbf{k}) - 2q \cos(2x)  \right] \tilde u_\textbf{k} = 0,
 \label{Eq:Mathieu_real}
 \end{equation}}

where $ A(\textbf{k}) = 1 + \dfrac{k^2}{\mu^2 \ a^2}$ is a rescaling of the wavenumber and $q = 3 \dfrac{H}{\mu} + \mathcal{O}^2\left(\dfrac{H}{\mu}\right)$ modulates the instability. On large scales (for instance, CMB scales), the conservation of curvature perturbation is sufficient to establish that the power spectrum calculated at the end of inflation propagates through the reheating epoch without being distorted. Since this real SF is in the fast oscillating phase, it can be guaranteed that $q \ll 1$, therefore, we are in the narrow resonance regime. We can determine the first instability band, that is given by,
\begin{equation}
1 - q < A_\textbf{k} < 1 + q,
\label{Eq:instability_math}
\end{equation}

and which can be expressed as, 
\begin{equation}
0 < k < a\sqrt{3H \mu}.
\label{Eq:instability_math2}
\end{equation}

The new characteristic spatial length given by $\ell_c \equiv 1/\sqrt{3H \mu}$, sets the lower bound of the instability band, the oscillations of the background give way to the resonance responsible for such instability.  

{The instability scale is depicted in Figure~\ref{Fig:jeans_band_scales} as Mathieu instability, which illustrates the difference with the instability of the complex SF.}

\Correx{Mathematically, the Mathieu equation and associated oscillations correspond to the angular part of the wave equation in cylindrical  coordinates. The radial oscillations can be represented through a slight modification of Eq.~\eqref{Eq:Mathieu_real}, with $x \to i x$ \cite{arscott1964}.}

\providecommand{\href}[2]{#2}\begingroup\raggedright\endgroup

\bibliographystyle{JHEP}

\end{document}